\documentclass[pra,%
 reprint, onecolumn,
 amsmath,amssymb,
 aps,
]{revtex4-2}
\usepackage{graphicx}
\usepackage{dcolumn}
\usepackage{bm}
\usepackage{hyperref}
\usepackage{url}

\usepackage{xcolor}
\usepackage{caption}
\usepackage{subcaption}
\usepackage{multirow}
\usepackage[dvipsnames]{xcolor}

\usepackage{enumitem,amssymb}
\newlist{todolist}{itemize}{2}
\setlist[todolist]{label=$\square$}
\usepackage{pifont}

\usepackage{amsmath}

\DeclareMathOperator{\sign}{sign}

\begin{document}

\preprint{APS/123-QED}

\title{Nonlinear Tearing Modes in Current-Vortex Sheets}

\author{D. Urbanski\footnote{Corresponding author. Email: deu223@utexas.edu}}
\author{F. L. Waelbroeck}
\author{A. Tenerani}%
\affiliation{Institute for Fusion Studies and Physics Department, University of Texas at Austin}

\begin{abstract}
The linear and nonlinear development of instabilities and Alfv\'en resonances in a plane current-vortex sheet is presented here for sheared equilibrium profiles $\boldsymbol{B_{y0}} = \tanh(z)\boldsymbol{\hat{y}}$ and $\boldsymbol{V_{y0}} = M_0\tanh(z/r)\boldsymbol{\hat{y}}$. We extend Rutherford's nonlinear model for constant-psi magnetic islands to account for a sheared equilibrium flow and determine the flow's impact on the magnetic island's nonlinear evolution for $M_0<1$. In this regime, the flow introduces two nonlinear contributions: a polarization current term and a modification of the saturation term derived by \citet{MiPo}. We find that the polarization current generated by the equilibrium flow consistently reduces the nonlinear growth rate of the tearing mode. The flow-induced modification of the saturation term, however, depends on the magnetic-to-velocity shear width ratio, $r$. For $r=1$, the saturation term is unchanged from the  case without flow. For $r>1$, it increases in magnitude with $M_0$, strengthening its stabilizing effect and further reducing the island growth rate. In contrast, for $r<1$, the saturation term decreases in magnitude with increasing $M_0$ and can become positive, leading to a transition from a stabilizing to a destabilizing contribution. Nevertheless, throughout the parameter range considered, the combined effect of the polarization current and modified saturation term is to slow the growth of the magnetic island. Finally, we find that, in the presence of Alfv\'en resonances, the magnetic island's growth in the nonlinear regime is no longer adequately characterized by constant-psi, and the dynamics of such islands are not captured by the model.
\end{abstract}

\maketitle

\section{Introduction}\label{sec:Intro}
In a fluid, the presence of a sheared equilibrium flow can lead to the development of the fast-growing, ideal Kelvin-Helmholtz instability (KHI) \cite{chandrasekhar1961hydrodynamic}. Analogously, a magnetized plasma with a sheared equilibrium magnetic field can be unstable to the resistive Tearing Mode (TM) instability that causes the formation of magnetic islands through reconnection \cite{FurthKR}. In many space and laboratory plasmas, velocity shear and magnetic shear coexist within the same structures, leading to a complex interplay between KHI and TM.

In the solar wind, current sheets embedded within sheared flows are observed across a wide range of scales. At large scales, the heliospheric current sheet often displays signatures of magnetic reconnection such as exhausts, energized particles, and plasmoids \cite{phan2021prevalence,phan2022parker,liewer2023structure,sanchez2019situ}. At smaller scales, current sheets naturally arise within Magnetohydrodynamic (MHD) turbulence, often embedded in turbulent eddies exhibiting both current and vortex sheets \cite{Boldyrev2018,Boldyrev2006,borgogno2022coexistence}, where  velocity gradients  may inhibit reconnection \citep{mallet2025effect}. 
Coupling between KHI and TM is also observed in planetary environments \cite{johnson2014kelvin}, and is believed to leave signatures in solar flares \cite{Parker2017}. At Earth’s magnetopause, velocity shear between the magnetosheath and magnetosphere drives KHI, which can induce magnetic reconnection by distorting field lines during vortex roll-up \cite{birn1992quasi,chen1997tearing,tenerani2011nonlinear}.

In magnetic confinement devices, magnetic shears dominate the flow over most of the plasma domain, and the field coupling is minimal. However, in regions where the magnetic shear reverses, such as the core of reversed-shear tokamaks, the flow can be relatively large and alter plasma dynamics \cite{Ebrahimi2008}. Another such region exists in the pedestal, caused by the large bootstrap current driven there \cite{AydemirPlasmaEdge}.

Investigating how sheared flows interact with current sheets and affect reconnection is therefore important for understanding energy transfer, turbulence, and plasma transport in both space and laboratory plasmas. In this work, we investigate reconnection in the presence of a sheared flow using the equilibrium profiles 
\begin{equation}
\boldsymbol{B_0} = \tanh(z)\boldsymbol{\hat{y}},
\end{equation}
\begin{equation}
\boldsymbol{V_{0}} = M_0\tanh(z/r)\boldsymbol{\hat{y}}, 
\end{equation}
where $M_0$ is the flow speed normalized to the Alfv\'en speed and $r$ is the ratio of the velocity shear width to the magnetic shear width. Although this configuration has been widely adopted, previous studies have primarily focused on specific regions of the parameter space $(M_0, r)$, as summarized below.

The case $r=1$, corresponding to identical velocity and magnetic shear widths, has been extensively studied. In this limit, sub-Alfv\'enic flows ($M_0 < 1$) lead to tearing instability with reduced growth rates, while super-Alfv\'enic flows ($M_0 > 1$) drive the KHI \cite{miura1982nonlocal,einaudi1986resistive,chen1997tearing,mallet2025effect}. The linear analysis for the case $r\neq1$ was carried out by \citet{chen1990resistive}, who derived the TM growth rate and demonstrated that the tearing index, $\Delta^\prime$, is significantly modified by the flow. Additionally, the dependence of the linear growth rate on viscosity was investigated by \citet{Dahlburg} and \citet{Coelho2007} in Cartesian and cylindrical coordinates, respectively. A nonlinear study was conducted by \citet{Dahlburg}, who characterized instabilities along the cuts in parameter space $(M_0,r=1)$ and $(M_0=1,r)$. Their study demonstrated that unstable modes are present for $r<1$. They also carried out a linear energy analysis and demonstrated that the presence of both an equilibrium current sheet and an equilibrium vortex sheet allows for new energy extraction from the fields into the instabilities.

Despite these efforts, the regime $r \neq 1$ remains comparatively unexplored, even though it is more representative of realistic systems and supports richer dynamics. In particular, there exist two domains in the parameter space spanned by $M_0$ and $r$ ($r < M_0 < 1$ and $1 < M_0 < r$) in which Alfv\'en resonances arise. These resonances typically develop in sheared magnetic fields in the presence of an external driver \cite{UrbanskiPaper1,uberoi1999alfven}, but they have also been identified in  simulations with tearing mode configurations. In cartesian coordinates, these resonances were studied in the nonlinear simulations of KHI \cite{Li2016} and in super-Alfv\'enic, compressible simulations of \citet{Li2012}. These resonances have also been identified in cylindrical geometry configurations by \citet{Wei2013} for the $m/n = 2/2$ and $3/3$ modes.  Their presence is of particular interest in the present context because they can extract energy from the reconnection layer, thereby suppressing tearing-driven island growth \cite{UrbanskiPaper1}, or modify island size \cite{Li2016}. However, their origin and role in systems with coupled velocity and magnetic shear have not been systematically investigated.

Moreover, most prior studies have focused on reconnection induced by KHI-driven vortex roll-up at large Mach numbers \cite{Knol02,Li2012,Li2016,Li2021}, leaving the complementary regime, in which tearing modes dominate in the presence of shear flows, largely unexplored in the nonlinear stage. As a result, a comprehensive understanding of how equilibrium flow shear modifies both the onset and the nonlinear evolution of tearing-driven magnetic islands is still lacking.

The goal of this work is therefore twofold. First, we characterize the linear stability properties of the system across the full parameter space $(M_0, r)$, identifying regimes dominated by KHI or TM, as well as regions where Alfv\'en resonances arise. Second, we investigate how equilibrium shear flows modify the nonlinear evolution of tearing-driven magnetic islands, with particular emphasis on their growth and saturation in the constant-psi regime.

This question is particularly relevant in magnetic confinement devices, where magnetic islands degrade confinement by modifying the magnetic topology and energy balance \cite{fitzpatrick2023theoretical, Chandra2015}. For this reason, there has been much interest in understanding the impact of flows on the neoclassical tearing mode \cite{Haye2010,Gerhardt2009,Lahafloshr,Buttery2008,White2015,ChandraSenNF05,ofman1993nonlinear}. More broadly, the size and evolution of magnetic islands play an important role in particle acceleration and cross-scale energy transfer in turbulent plasmas \cite{drake2006electron,dong2018role}. In the absence of flow, the nonlinear evolution of the tearing mode is well described by Rutherford theory in the constant-psi regime \cite{Rutherford}. Building on the seminal work of \citet{Rutherford} and its extension to include shear flows in the fully reconnected state \cite{waelbroeck2007effect, waelbroeck1997rotation}, we develop a theoretical framework to describe the nonlinear evolution of tearing-driven magnetic islands in the presence of an equilibrium shear flow to understand how flow modifies island growth and saturation.

This paper is organized as follows. The model equations are presented in section \ref{sec:Model Eqs}. In section \ref{sec:Linear}, we identify regions where KHI, TM, and Alfv\'en resonances exist in the parameter space ($M_0, r)$. In section \ref{sec:Nonlinear}, we consider the TM-dominated regime and derive a nonlinear model for the magnetic island width in the presence of equilibrium shear flow. Although we focus our efforts on the TM-dominated regime, our results are also applicable to the case of the saturated KHI, as current sheets may coexist with vortex sheets in the relaxed state after roll-up. The model is compared with simulations in section \ref{sec:Numerical Simulations}. Concluding remarks are made in section \ref{subsec:Conclusions}.

\section{Model Equations} \label{sec:Model Eqs}
In this study, we consider a two dimensional plasma in the $y-z$ plane normal to $\hat{x}$. Denoting $\perp$ as the directions perpendicular to the normal vector, the in-plane magnetic field is defined as $\boldsymbol{B_\perp} = \hat{x}\times\nabla\psi$ and $\boldsymbol{V_\perp} = \hat{x}\times\nabla\phi$, where  $\psi$ and $\phi$ are the flux and stream functions. The reduced MHD (RMHD) model is adopted, which is described by the following equations,
\begin{equation} \label{eq:Induction Equation}
\frac{\partial \psi}{\partial t} = - [\phi,\psi] +S^{-1} (J-J_0),
\end{equation}
\begin{equation}  \label{eq:Equation of Motion}
\frac{\partial U}{\partial t} = - [\phi,U] - [J,\psi] + R^{-1}\nabla^2 (U - U_0),
\end{equation}
where $J={\hat{x}}\cdot{\boldsymbol {J}} = \boldsymbol  \nabla_\perp^2\psi $ and $U = \boldsymbol {\hat{x}}\cdot {\boldsymbol {U}} = \nabla_\perp^2\phi$ are the out-of-plane current density and vorticity, respectively. The reference state current and vortex sheets are maintained by external sources. This is accounted for with the subtraction of the $J_0$ and $U_0$ in the diffusive terms, and also means that $\frac{\partial\psi_0}{\partial t} = 0$ and $\frac{\partial U 0}{\partial t} = 0$. Here, $\left[f,g\right] = \partial_yf\partial_zg - \partial_zf\partial_yg$ is the Poisson bracket. The magnetic field has been normalized to $B_0$, the magnitude of the magnetic field far from the current sheet. All length scales have been normalized to the width of the equilibrium current sheet, $a$, all velocities to the Alfv\'en speed, $v_a =  B_0 / \sqrt{4\pi \rho_0}$, and time scales to the Alfv\'en crossing time, $\tau_a = a/v_a$. The plasma density $\rho_0$ is assumed to be uniform. $S$ and $R$ are the Lundquist and Reynolds numbers measuring the ratios of the resistive and viscous diffusive times scales, $\tau_R = \frac{a^2}{\eta}$ and $\tau_\nu = \frac{a^2}{\nu}$, to the Alfv\'en crossing time.

The unperturbed equilibrium is defined by the following profiles for $\boldsymbol{B_\perp}$ and $\boldsymbol{V_\perp}$,
\begin{equation}\label{eq:B0}
\boldsymbol{B_{y0}} = \tanh(z)\boldsymbol{\hat{y}};
\end{equation}
\begin{equation}\label{eq:V0}
\boldsymbol{V_{y0}} = M_0\tanh(z/r)\boldsymbol{\hat{y}},
\end{equation}
where $M_0$ is the flow speed normalized to the Alfv\'en speed and $r$ is the ratio of the vortex sheet width to the current sheet  width. We assume that $\phi$ and $\psi$ vanish as $|z| \to \infty$ and impose periodic boundary conditions in the $y$ direction with associated wave number $k$.

\section{Linear Analysis: Marginal Stability of Ideal and Resistive Modes in the Presence of Sheared Flows}\label{sec:Linear}

In the linear approximation, the RMHD equations \eqref{eq:Induction Equation}-\eqref{eq:Equation of Motion} define a system characterized by the parameter space $(M_0, r, k, S, R)$. To guide the choice of parameters for the nonlinear analysis, we explored this parameter space and identified distinct domains: the instability domains for the Kelvin-Helmholtz mode associated with strong velocity shear and for the tearing mode, which drives magnetic reconnection at the resonant surface ${\bf k\cdot B}_0=0$. The nature of the modes can be affected by Alfvn resonances that arise where the local phase velocity matches the Alfv\'en speed. Although we seek to study the impact of sheared flow on the nonlinear growth of the tearing mode, which is resistive by nature, the ideal RMHD model equations ($S = \infty$, $R=\infty$) are enough to determine the stability of the ideal mode and the tearing mode parameter, $\Delta^\prime$, defined below, which measures the stability of the tearing mode in the absence of flow.

We begin by linearizing Eqs.~\eqref{eq:Induction Equation} and~\eqref{eq:Equation of Motion} about the equilibrium defined in Eqs.~\eqref{eq:B0} and~\eqref{eq:V0}, retaining terms up to first order in the perturbations. We assume perturbations of the form $f_1 = A(z)\exp(iky - i\omega t)$. Due to the parity of the equilibrium, the real component of the frequency, $\omega$, is zero and $\omega = i\gamma$, where the imaginary component, $\gamma$, measures the linear growth of the instability \cite{chandrasekhar1961hydrodynamic}.
The linearized equations are
\begin{equation}\label{eq:Linear Induction}
-(\frac{i\gamma}{k}-V_{y0})\psi_1 + (i/Sk)(\psi_1^{\prime\prime}-k^2\psi_1)=B_{y0}\phi_1,
\end{equation}
\begin{equation}\label{eq:Linear Equation of Motion}
B_{y0}(\psi_1^{\prime\prime}-k^2\psi_1)-B_{y0}^{\prime\prime}\psi_1=-(\frac{i\gamma}{k}-V_{y0})(\phi_1^{\prime\prime}-k^2\phi_1) -V_{y0}^{\prime\prime}\phi_1+ (i/Rk) (\phi_1^{(4)} - 2 k^2\phi_1^{\prime\prime}+k^4\phi_1),
\end{equation}
where the prime denotes the derivative with respect to $z$. In an ideal plasma for which $S=\infty$ and $R = \infty$, Eq.~\eqref{eq:Linear Induction} becomes
\begin{equation}\label{eq:Ideal Linear Induction}
\psi_1=-\frac{B_{y0}}{(\frac{i\gamma}{k}-V_{y0})}\phi_1.
\end{equation}
The electrostatic potential is proportional to the plasma displacement, so that the above equation reflects the advection of the flux by the plasma.
The above result allows the vorticity equation \eqref{eq:Linear Equation of Motion} to be reduced to a single second-order differential equation,
\begin{equation}\label{eq:Alfven Resonances}
\frac{d}{dz}\left[\left[\left(\frac{i\gamma}{k}-V_{y0}\right)^2-B_{y0}^2\right]\frac{d\xi}{dz}\right] -k^2\left[\left(\frac{i\gamma}{k}-V_{y0}\right)^2-B_{y0}^2\right]\xi = 0,
\end{equation}
where $\xi$ is the transverse displacement
\begin{equation}\label{eq:Transverse Displacement}
\xi = -\frac{\psi_1}{B_{y0}}.
\end{equation}

The solution to Eq.~\eqref{eq:Alfven Resonances} has a singularity where the Doppler shifted phase velocity, $i\gamma/k - V_{y0}$, is equal to the local Alfv\'en speed, $\pm B_{y0}$ \cite{UrbanskiPaper1,einaudi1986resistive,mok1985resistive}. With our choice of equilibrium, for which ${\bf B}{y0}$ and ${\bf V}{y0}$ are antisymmetric, it is the positive solution, $(\frac{i\gamma}{k}-V_{y0})+B_{y0}$, which is resonant on both sides of the magnetic neutral line. It is possible to identify a clear domain in the parameter space in which the Alfv\'en resonances are present. For large $z$, the equilibrium fields are asymptotically flat, and for $M_0 = 1$, the fields overlap for all values of $r$, conceptually producing a resonant surface. This constitutes an upper limit of the phase space parameters to which resonance can occur, with the resonances appearing outside of both the current and vortex sheets. For slowly time-varying fields away from the tearing layer, $\gamma \rightarrow0$, and the condition for resonances is $V_{y0} = B_{y0}$, or in this case $M_0\tanh(z/r) = \tanh(z)$. In the limit of small z, this condition approaches $M_0/r = 1$, or $M_0 = r$. This is the lower limit for resonances, with resonances approaching the neutral line of the equilibrium current sheet.

We define four regions organized by the presence of Alfv\'en resonances. In region I, $1 < M_0$ and $r < M_0$, and the flow is everywhere super-Alfv\'enic. In region III, $M_0 < 1$ and $M_0 < r$ and the flow is everywhere sub-Alfv\'enic. As a consequence, neither region is resonant. In region II, for which $1 < M_0 < r$, the flow is super-Alfv\'enic everywhere except between the Alfv\'en resonances that enclose the neutral plane $z=0$. In this region, the effect of increasing $M_0$ is to generate resonances closer to the neutral line. Conversely, in region IV, where $r<M_0<1$, the flow is sub-Alfv\'enic outside the Alfv\'en resonances but super-Alfv\'enic in between. The resonances here are generated farther from the neutral line with increasing $M_0$.

We now numerically determine the stability conditions for the ideal mode in the parameter space spanned by ($M_0$, $r$, $k$). In the absence of an equilibrium flow, the imaginary and real components of the potentials are decoupled, and the system is treated as real. With the inclusion of flow, the real and imaginary components cannot be trivially separated, and the system to solve is a set of four ordinary differential equations spanned by $\phi_r = \Re{[\phi_1]}$ and $\phi_i = \Im{[\phi_1]}$ and their first derivatives.

\begin{figure}
\centering
\includegraphics[width=0.5\linewidth]{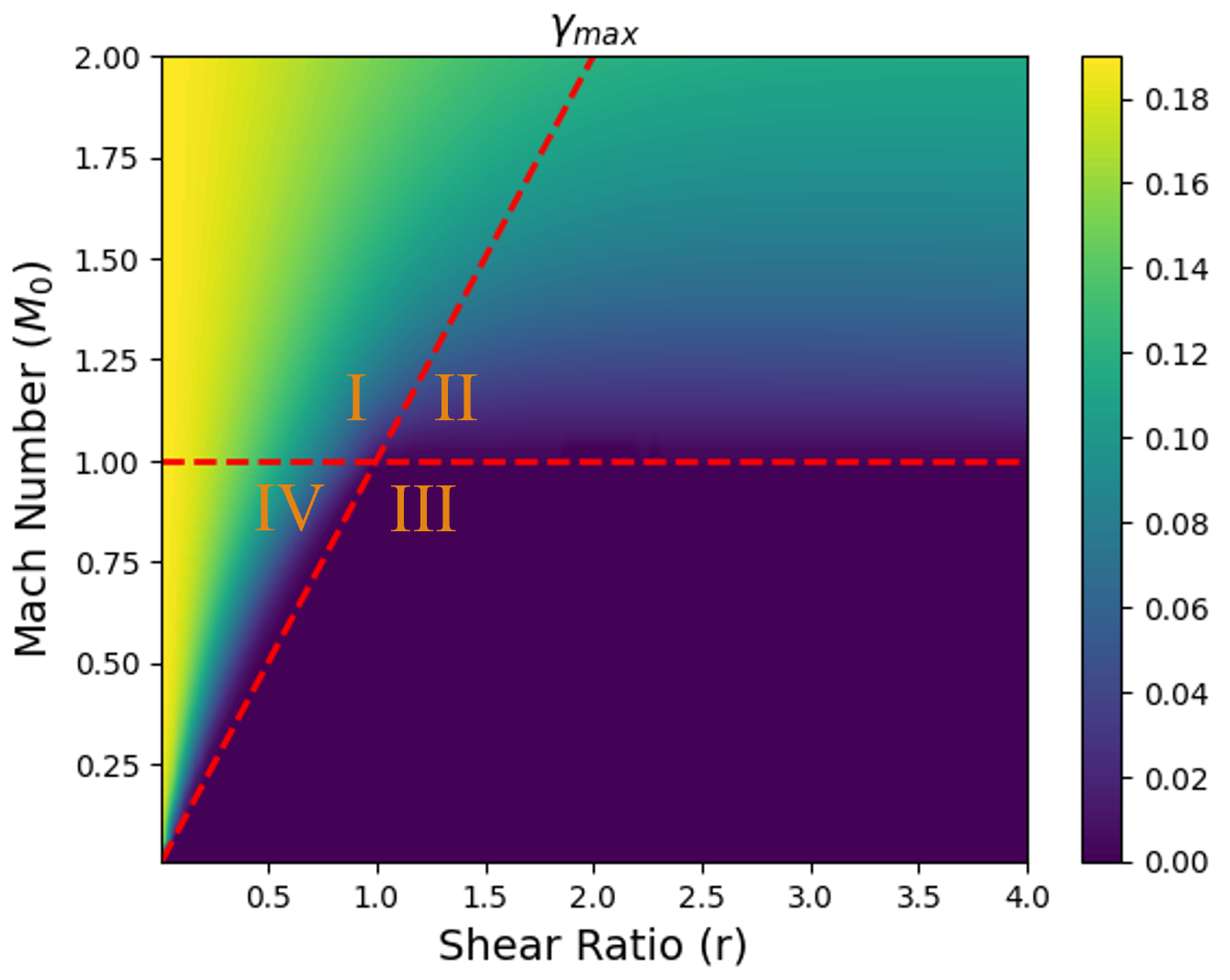}
\caption{The dispersion relation of the ideal instability, which relates the growth rate to the mode wave number $k$, is calculated numerically for various pairs of the Mach number $M_0$ and the ratio of flow to magnetic shear widths $r$. The maximum growth rate, corresponding to a wavenumber $k_{max}$, is determined for each pair choice of $M_0$ and $r$, and a contour plot of the maximum growth rate of the ideal instability as a function of $M_0$ and $r$ is plotted here. The dashed red lines indicate the $M_0=r$ and $M_0 = 1$ asymptotic limits, which define a region in which Alfv\'en resonances are present. Note that, to improve the plot's readability, the growth rate here is normalized to the flow time scale, $\tau_v = r/M_0$, and differs from the rest of the paper, in which quantities are normalized to the Alfv\'en time scale.}
\label{fig:Ideal Parameter Space}
\end{figure}

We solve in the parameter space spanned by $M_0$, $r$, and $k$ with a shooting method code, assuming even parity in $\phi_r$ and using as shooting variables $\phi_i ^\prime (z=0)$ and $\gamma$. With two shooting variables, we require two boundary conditions at $z = z_b$, for which we chose the ansatz $\phi_i(z_b) = \exp{(-k|z_b|)}$ and $\phi_r^\prime(z_b) =0$.  The boundary condition, $\phi_i(z_b) = \exp{(-k|z_b|)}$, is the solution to Eqs.~\eqref{eq:Ideal Linear Induction}-\eqref{eq:Transverse Displacement} for $z\rightarrow\infty$.  At each value of $M_0$ and $r$, the dispersion relation, which relates the growth rate of the ideal instability $\gamma$ to the wavenumber $k$, is calculated numerically. The maximum growth rate of the dispersion relation for each choice of parameter pair ($M_0$, $r$) is plotted as a contour plot spanning this parameter space in Fig.~\ref{fig:Ideal Parameter Space}. Note that, to improve the plot's readability, the growth rate here is normalized to the flow time scale, $\tau_v = r/M_0$, and differs from the rest of the paper, in which quantities are normalized to the Alfv\'en time scale. We find that the KHI is stable when $M_0 < 1$ and $M_0 < r$, which is region III of the parameter space, where the equilibrium flow is everywhere sub-Alfv\'enic. Conversely, the plasma is unstable in regions I, II, and IV, where the flow is super-Alfv\'enic somewhere in the domain. These results indicate that a super-Alfv\'enic region is a necessary condition for the KHI to exist in this geometry to overcome magnetic tension forces. As far as we are aware, the presence of the ideal instability in region IV, parameterized by $r < M_0 < 1$, has not been previously reported.

When the plasma is unstable to the tearing mode, the effect of resistivity must be considered in a thin layer at the magnetic neutral line, defined by $\mathbf{k}\cdot \mathbf{B_0} = 0$. Resistive boundary layers also form at the Alfv\'en resonance sites and may lead to resistive instabilities in the absence of an ideal mode. Away from these boundary layers, however, the plasma is in a relatively steady state and is effectively ideal. This region is referred to as the outer region of the plasma. The series solution for the flux function in the outer region, governed by Eqs.~\eqref{eq:Alfven Resonances} and~\eqref{eq:Transverse Displacement}, asymptotically close to the thin, resistive layer, formally known as the inner region, is
\begin{equation}\label{eq:Psi Outer}
\tilde\psi_{out}(y,z) = \tilde\psi_{out}(y,z=0)\left[1 + \Delta^\prime\frac{|z|}{2} + \left(k^2 -
2 - 2\mu\right)\frac{z^2}{2}\right] + O(|z|^3),
\end{equation}
where
\begin{equation}\label{eq:mu}
\mu=\frac{2}{3}\frac{M_0^2}{r^2}\frac{(r^2-1)}{(r^2-M_0^2)}.
\end{equation}
Even parity is imposed as is appropriate for the tearing mode. Terms of order $z^2$ are kept in the expansion. \citet{MiPo} demonstrated that although these are small contributions to $\psi$, their derivatives can be large, and they provide an important contribution to the current density, $J$. It is interesting to note how one of the parameter space boundaries for resonances emerges in this expansion, $r^2-M_0^2$. Also, note how $\mu$ vanishes for the case $r=1$. This is consistent with past literature that has shown the outer region is unmodified by the presence of the flow when $r=1$ \cite{mallet2025effect}.

Here $\Delta^\prime$ is the tearing mode parameter, defined by the logarithmic discontinuity in the outer solution across the inner region, \begin{equation}\label{eq:Delta’}
\Delta^\prime = \frac{\psi_{\epsilon\rightarrow0_+}^\prime-\psi_{\epsilon\rightarrow0_-}^\prime}{\psi_{\epsilon\rightarrow0}}.
\end{equation}
\citet{chen1990resistive} showed with their equation 48 that the tearing mode parameter remains the appropriate stability parameter for our reference state, Eqs.~\eqref{eq:B0} and~\eqref{eq:V0}. Determination of $\Delta^\prime$ requires knowledge of the full solution of $\psi_1$ in the outer region, which can be found numerically with an ODE solver or from a simulation. For the case $r=1$, the outer solution is invariant to the flow amplitude, implying $\Delta^\prime$ is unmodified, which is consistent with past literature. The results of \citet{chen1990resistive} thus imply that the tearing mode is unstable in region III and region II, which together constitute $M_0 < r$.

We conclude that we can divide the parameter space spanned by $M_0$ and $r$ into four regions defined by the presence of Alfv\'en resonances and categorize the processes as follows
\begin{itemize}
\item  \textbf{Region I}  \textbf{($\mathbf{1 < M_0}$, $ \mathbf{r < M_0}$)}: the flow is everywhere super-Alfv\'enic and KHI is allowed.

\item  \textbf{Region II} \textbf{($\mathbf{ 1 < M_0 < r}$)}: the flow is asymptotically super-Alfv\'enic and resonances form at the boundary separating the super and sub-Alfv\'enic regions. KHI and TM are allowed.

\item  \textbf{Region III} \textbf{($\mathbf{M_0 < 1}$, $ \mathbf{M_0 < r}$)}: the flow is everywhere sub-Alfv\'enic and TM is allowed.

\item  \textbf{Region IV} \textbf{( $\mathbf{ r < M_0 < 1}$)}: the flow is asymptotically sub-Alfv\'enic with a super-Alfv\'enic region lying between the Alfv\'en resonances. KHI is allowed.

\end{itemize}
By ``allowed" we mean the presence of the instabilities depends on the wave number $k$, and that the flow amplitude alone isn’t a sufficient condition for the instabilities.

The defining feature of the tearing mode is the formation of magnetic islands due to magnetic reconnection at the magnetic neutral of the current sheet. The size of such islands, $W$, together with $\Delta^\prime$ characterize the dynamics of the tearing mode. While the magnetic island’s size is smaller than the linear layer width, $\delta \sim S^{-2/5}$, the magnetic flux, $\psi$, at the resonant surface (the neutral line) grows exponentially in time. The tearing mode enters the relatively slower, nonlinear stage when the island exceeds this size. For a sufficiently slow-growing magnetic island, parameterized by $\Delta^\prime W \ll 1$, resistive diffusivity smooths out non-uniformities such that the magnetic flux is roughly constant in the island. This is known as the constant-psi regime, and an island which violates its assumptions is known as a non-constant-psi magnetic island.

In the absence of flow, \citet{Rutherford} developed a model which describes the behavior of magnetic islands which enter the nonlinear stage while in the constant-psi regime. In the following sections, we extend Rutherford’s model to include the effects of an Alfv\'enic flow on magnetic islands generated by TM in regions III and IV. In region IV, we test the model for cases in which the KHI is stable by selecting an appropriate minimum periodic wavenumber $k$ to understand the impact of the Alfv\'en resonances on the growth of the magnetic island.

\section{Nonlinear theory of The Effect of Sheared Flow on Magnetic Island Evolution}\label{sec:Nonlinear}

We take an asymptotic matching approach to determining the contributions of the Alfv\'enic flow to the magnetic island size. Solutions to the inner and outer regions, within and far from the resonant layer, respectively, are to be matched in the intermediate region defined by the width of the magnetic island and the equilibrium, $W \ll z \ll \min(1,r)$. The upper limit conditions that the island be smaller than both the width of the current and the vortex sheets. The solution in the outer region to be matched to the inner region was provided in section \ref{sec:Linear} by Eq.~\eqref{eq:Psi Outer}. We now seek to match it to the inner region.

In the inner region, gradients are strong, $\partial_z^2 \gg \partial_y^2$, and resistivity plays a significant role. If $\delta$ is the layer width, it is assumed that the magnetic island grows on a time scale much longer than the resistive diffusion time for the layer, $\delta^2/\eta$, so that variations in $\psi$ are smoothed out by resistive diffusion. This is known as the constant-psi approximation and is otherwise stated as
\begin{equation}
    \psi(y,z,t) = \psi_{0}(z) + \tilde{\psi}(y,t).
\end{equation}
Expressing the perturbation as a Fourier series, we have
\begin{equation}\label{eq:psi-tilde}
    \tilde{\psi}(Y,t) = \sum_{n=1,\infty}\tilde{\psi}_n(t)\cos(n Y),
\end{equation}
where $Y = k y$. Near marginal stability for the tearing mode, harmonics higher than the fundamental do not contribute much to the shape of the island, so we keep only the fundamental harmonic
\begin{equation}\label{eq:Psi Constant-psi}
    \psi(Y,z,t) = \psi_0(z) + \tilde{\psi}_1(t)\cos(Y).
\end{equation}
Taylor expansion of the equilibrium flux function leads to the familiar form for constant-psi magnetic flux,
\begin{equation}\label{eq:Psi Constant-psi Expansion}
    \psi(Y,z,t) = -\frac{z^2}{2} + \tilde{\psi}_1(t)\cos(Y).
\end{equation}
The magnetic island size is defined by
\begin{equation}\label{eq:Island Size}
    W = 4\sqrt{\tilde{\psi}_1}.
\end{equation}
By requiring that the outer solution match with the inner layer solution described by constant-psi, \eqref{eq:Psi Constant-psi}, at the origin, it is found that
\begin{equation}\label{eq:psi out Amp}
    \tilde\psi_{out}(0) = \tilde{\psi}_1 \cos(Y).
\end{equation}
In the outer region, two expansions were undergone: a linear expansion about the perturbations of the potentials, $\psi_1$ and $\phi_1$, and a small $z$ $(z\rightarrow0)$ expansion in order to match to the inner region. In the inner region, there are two appropriate expansions: we expand for large $z$ $(z\rightarrow\infty)$ for the matching, but also about the magnetic island width, $W \sim \tilde\psi_1^{1/2}$, which is small by virtue of the condition that the island be smaller than the widths of both the current and vortex sheets, $W \ll \min(1,r)$. The smallness of terms is thus assessed by the relative powers of $W$.

We seek smoothness of $\psi$ in the matching region of the inner and outer regions. This is imposed by requiring its first derivative to be continuous at the boundary. Integration of the out-of-plane current density $J = \boldsymbol\nabla^2_\perp\psi$ provides the matching condition, and the logarithmic discontinuity of the outer solution is the boundary condition for the inner solution
\begin{equation}\label{eq:Matching}
    \oint dY
\left(\frac{\partial\tilde\psi_{out}}{\partial z}\bigg\vert^{z_
\epsilon}_{-z_\epsilon}+2z_{\epsilon}\frac{\partial^2\tilde\psi_{out}}{\partial y^2}\right)\cos(Y)  =  \oint dY \int^{z_
\epsilon}_{-z_\epsilon}dz J \cos(Y).
\end{equation}
The position $z_\epsilon$ is an arbitrary position for matching inside the matching region $W \ll z_\epsilon \ll \min(1,r)$. On the left hand side, which is the jump in the outer solution asymptotically close to the inner layer, integration of the equilibrium state current density, $J_0$, vanishes since there it exhibits no jump. $\tilde{\psi}(Y,t)$ and $J$ are each a series of nonlinear harmonics, defined by Eq.~\eqref{eq:psi-tilde}. We select out the $n=1$ harmonic explicitly by multiplying each side by $\cos(Y)$ before integration. Upon integration, the factor of $\cos(Y)$ has the effect of projecting $J$ along the $n=1$ harmonic, as integration of harmonics $m \neq n$ will add negligible contributions \cite{Fitzpatrick}. Harmonics with different order, with $n \neq 1$, can be found by multiplying instead with $\cos(nY)$ and performing the same integration.

We will now focus our efforts on determining the current density in the inner region to be matched in \eqref{eq:Matching}. The inclusion of an equilibrium flow modifies the ordering of terms in the RMHD model compared to the traditional case without flow. For a slowly varying plasma in the presence of an equilibrium flow, Ohm's law, Eq.~\eqref{eq:Induction Equation}, is to lowest order in $\eta$ ($\sim S^{-1}$)
\begin{equation}\label{eq:Induction Steady}
    -[\phi,\psi] = \mathbf{B}\cdot\nabla\phi = 0.
  \end{equation}
Here, we mean slowly varying relative to the Alf\'en wave crossing time, as is consistent with the constant-psi ordering. This means the stream function, $\phi$, which is the electrostatic potential, is constant along magnetic field lines. In other words, the stream function is constant along surfaces of constant magnetic flux, $\psi$, as the magnetic field lines are the contours of $\psi$. Upon integration of $\psi$ this tells us
\begin{equation}\label{eq:Phi Profile}
    \phi =  \Phi_\sigma(\psi) + O(\eta),
\end{equation}
with $\sigma$ labeling the branches of the profile function $\Phi$ lying on different sides of the magnetic island separatrix. In this form, we are transitioning from a coordinate system defined by $(z, y)$ to one defined by $(\psi, y)$ in which $z$ is a function of the new coordinates. In this paper, we will assume the flux surfaces are symmetric across the separatrix and drop the subscript $\sigma$.

In the absence of the flow, the equation of motion, Eq.~\eqref{eq:Equation of Motion}, in the steady state is to first order $[\psi,J] = 0$, which informs us that the current density is constant along magnetic flux surfaces. This is not true when flow is present, and the ordering is changed. In the presence of an equilibrium flow, the steady state of the vorticity equation, Eq.~\eqref{eq:Equation of Motion}, is to lowest order
\begin{equation}\label{eq:Equation of Motion Steady}
    \left[ \phi, U\right] - \left[ \psi, J\right] = 0,
\end{equation}
after minor rearrangement of terms. Substitution of \eqref{eq:Phi Profile} into \eqref{eq:Equation of Motion Steady} with the property $[\Phi(\psi),U] = [\psi,\Phi'(\psi)U]$, provides the integratable function
\begin{equation}\label{eq:Equation of Motion Steady psi}
    \left[ \psi, \Phi'(\psi)U-J\right] = 0,
\end{equation}
which has the solution
\begin{equation}\label{eq:J(I,U)}
    J = - I(\psi) + M(\psi)U,
\end{equation}
where $M(\psi) = \frac{d\Phi(\psi)}{d\psi}$ is the Alfv\'enic Mach number and $I(\psi)$ is another profile function describing the current profile. In the steady state, these profiles are determined by the saturation processes of the plasma. The out-of-plane vorticity, $U=\boldsymbol\nabla^2_\perp\Phi$, can be similarly expressed as a function of the profile functions,
\begin{equation}\label{eq:U profiles}
    U = JM(\psi) + B_{\perp}^2 M'(\psi),
\end{equation}
in which the prime denotes a derivative with respect to $\psi$. Upon replacement in \eqref{eq:J(I,U)}, we find a form of the current density which is a function of the profile functions and the magnetic field \cite{waelbroeck2007effect}
\begin{equation}\label{eq:J(I,B)}
    J = \frac{-I(\psi)+B_{\perp}^2 M(\psi) M'(\psi)}{1-M^2(\psi)}.
\end{equation}
Together, Eqs.~\eqref{eq:U profiles} and~\eqref{eq:J(I,B)} specify an equilibrium state. It is observed that the profile functions, $I(\psi)$ and $M(\psi)$, determine the equilibrium configuration, which is formed by the physical processes of the plasma. 

We aim to determine the extent to which shear flow influences magnetic island size. The effect of flow in Eq.~\eqref{eq:J(I,B)} enters through the profile function, $M(\psi)$. Adapting Rutherford's approach, we derive a new term which quantifies this effect. To the next order in $\eta$, there is a contribution from the Poisson bracket, $[\phi,\psi]$, to Ohm's  law. In order to kill the Poisson bracket, we follow Rutherford's lead and define
the flux-surface average operator, which integrates the term to zero,
\begin{equation}\label{eq:Flux Surface Average}
    \langle f \rangle  = \frac{1}{2\pi}\int_s\frac{f}{B_y}kdy
\end{equation}
This integral is evaluated at constant values of $\psi$, or along magnetic flux surfaces, and is evaluated with different bounds inside and outside of the magnetic island
\[ \begin{cases}
      \langle f\rangle = \frac{1}{2\pi}\int^{2\pi}_{0} \frac{f}{B_y} dY & \Omega = \psi/\psi_{sep} > 1,\\
      \langle f\rangle =\frac{1}{2\pi} \int^{2\pi-Y_{tp}}_{Y_{tp}} \frac{f}{B_y} dY & \Omega < 1,\\
   \end{cases}
\]
where $Y = ky$, and $Y_{tp}$ is the turning point of magnetic field lines inside the island.

This procedure is consistent with other related literature. Note that this definition deviates from Rutherford's by the factor of $B_y^{-1}$ in the integrand. His operator was an average over the coordinate $y$, and his integrals are comparable to ours as the functions he averaged contained factors of $B_y^{-1}$ in their denominators. Our definition is consistent with recent literature, simplifies upcoming equations, and in some cases results in integrals with known solutions. We take the flux surface average of both sides of Ohm's law, Eq.~\eqref{eq:Induction Equation}, an operation designed to eliminate the Poisson bracket. The remaining terms are, to next order in $\eta$ ($\sim S^{-1}$),
\begin{equation}\label{eq:Flux Integral Induction}
    \langle\frac{\partial\psi}{\partial t}\rangle =  S^{-1}\big(\langle J \rangle - \langle J_0 \rangle \big).
\end{equation}

Ultimately, we seek a form of $J$ in the vicinity of the magnetic neutral line to be matched asymptotically with a steady state, outer solution. Doing so will allow us to model the nonlinear evolution of a constant-psi magnetic island in the presence of a vortex sheet. In Eq.~\eqref{eq:J(I,B)}, the current density depends on the arbitrary profiles $I(\psi)$ and $M(\psi)$. With the manipulation of Eqs.~\eqref{eq:J(I,B)} and~\eqref{eq:Flux Integral Induction}, we find an expression for $J$ which depends only on the profile $M(\psi)$, which measures the effect of flow. Substitution of \eqref{eq:J(I,B)} into \eqref{eq:Flux Integral Induction} results in
\begin{equation}
    \langle\frac{\partial\psi}{\partial t}\rangle  + S^{-1}\langle J_0 \rangle= S^{-1} \left( -\frac{I(\psi)\langle1\rangle}{1-M(\psi)^2}+\frac{M(\psi)M(\psi)' \langle B_\perp^2 \rangle}{1-M(\psi)^2}\right),
\end{equation}
where we have used the fact that $I(\psi)$ and $M(\psi)$ are independent of the coordinate $y$ and pulled them outside of the flux surface average operators. Solving for $I(\psi)$,
\begin{equation}\label{eq:I}
    I(\psi) = -\left(1-M(\psi)^2\right)\frac{S\langle\frac{\partial\psi}{\partial t}\rangle + \langle J_0 \rangle}{\langle1\rangle} + M(\psi) M'(\psi)\frac{\langle B_\perp^2 \rangle}{\langle 1 \rangle},
\end{equation}
and substituting back into Eq.~\eqref{eq:J(I,B)}, we find the current density in the layer to be asymptotically matched in Eq.~\eqref{eq:Matching},
\begin{equation}\label{eq:J(M)}
    J = \frac{S\langle\frac{\partial\psi}{\partial t}\rangle + \langle J_0\rangle}{\langle1\rangle} + \frac{M(\psi) M'(\psi)}{(1-M^2(\psi))}\left(B_{\perp}^2-\frac{\langle B_{\perp}^2\rangle}{\langle1\rangle} \right).
\end{equation}

Substituting our asymptotic solution to $\psi$ in the outer region, Eq.~\eqref{eq:Psi Outer}, and the inner layer solution for $J$, Eq.~\eqref{eq:J(M)}, into the matching condition \eqref{eq:Matching} provides a model which generalizes the Rutherford to include the effects of Alfv\'enic flow,
\begin{equation}\label{eq:Matched J0 Symmetric}
    \begin{split}
    \Delta' - 4(1+\mu)z_{\epsilon} &= \frac{4S}{ \tilde{\psi}_1^{1/2}} \frac{d\tilde{\psi}_1}{dt} A
    +\frac{4}{ \tilde{\psi}_1^{1/2}} \int^{1}_{\Omega_-}d\Omega\left< \frac{\langle J_0\rangle_\Omega}{\langle1\rangle_\Omega} \cos(Y)\right>_\Omega
    \\& + \frac{4}{\tilde{\psi}_1^{3/2}} \int^{1}_{\Omega_-}d\Omega\frac{M(\tilde\psi_1\Omega) M'(\tilde\psi_1\Omega)}{(1-M^2(\tilde\psi_1\Omega))}\left<\left(B_{\perp}^2-\frac{\langle B_{\perp}^2\rangle_\Omega}{\langle1\rangle_\Omega} \right)\cos(Y)\right>_\Omega,
\end{split}
\end{equation}
where $A$ is Rutherford's integral,
\begin{equation} \label{eq: Rutherford A}
    A = \int^{1}_{\Omega_-}d\Omega\frac{\langle\cos(Y)\rangle^2_\Omega}{\langle1\rangle_\Omega}.
\end{equation}
which in the limit $\Omega_- \rightarrow \infty$ equals 0.41. The intermediate steps for this derivation are provided in the appendix, but the key steps are summarized here. The constant $\psi$ approximation has been assumed such that
\begin{equation}\label{eq:By constant-psi}
    B_y = -\frac{d\psi}{dz}\approx \sign(z)\sqrt{-2(\psi - \tilde{\psi}_1\cos(Y))},
\end{equation}
and $\psi$ everywhere has been normalized to $\tilde \psi_1$ to make the island width dependence more transparent. This includes the flux surface averages, which have been normalized by $\langle f \rangle = \tilde\psi_1^{-1/2}\langle f\rangle_{\Omega}$ and the derivative in $M^\prime$, which is now a derivative with respect to $\Omega$. Symmetry is assumed across the magnetic neutral line, and a factor of 2 is picked up. Finally, for the equilibrium profiles imposed, Eqs.~\eqref{eq:B0} and~\eqref{eq:V0}, $\Omega_{\pm}$ is negative. For this reason, we have flipped the integral in $\Omega$ to integrate from a negative boundary condition to the positive boundary condition provided at $z=0$, $\Omega = 1$, and pick up a factor of -1. As a consequence, the factor $\sign(z)$ in \eqref{eq:By constant-psi} is positive.

We seek a model which captures, to the lowest order in $W \sim \tilde \psi_1^{1/2}$, the effect of the Alfv\'enic flow on the growth of the magnetic island. To lowest order the flux function $\psi$ goes as $W^2$ (or $z^2$), according to Eq.~\eqref{eq:Psi Constant-psi}, but \citet{MiPo} showed that although next order contributions to the flux function are of order $O(W^4)$, their derivatives can lead to sizable contributions for small $\Delta^\prime$, as is the case when the tearing mode is close to marginal stability. This is important, particularly when considering the contribution of the current density, which goes as twice the derivative of $\psi$ in $z$. The equilibrium current is, correct to order $O(W^2)$, 
\begin{equation}\label{eq:J0 Equilibrium}
    J_0 = -1 + z^2 + O(z^4),
\end{equation}
and taken together with the constant-psi ordering, Eq.~\eqref{eq:Psi Constant-psi Expansion}, expansion of the last two terms leads to
\begin{equation}\label{eq:Matched J, Mach By exp}
    \begin{split}
    \Delta' - 4\mu z_{\epsilon} &= \frac{4S}{ \tilde{\psi}_1^{1/2}} \frac{d\tilde{\psi}_1}{dt} A 
    +8 \tilde{\psi}_1^{1/2} A
    \\& + \frac{8}{\tilde \psi_1^{1/2}} \int^{1}_{\Omega_-}d\Omega\frac{M(\tilde\psi_1\Omega) M'(\tilde\psi_1\Omega)}{(1-M^2(\tilde\psi_1\Omega))}\left( \langle\cos^2(Y)\rangle_\Omega-\frac{\langle\cos(Y)\rangle_\Omega^2}{\langle1\rangle_\Omega}\right).
\end{split}
\end{equation}

So far we have not assumed a form of $M(\tilde \psi_1 \Omega)$, beyond its definition $M = \frac{V_y}{B_y}$. To carry on, we assume $M$ to be equivalent to its equilibrium value, $M_{eq} = \frac{V_{y0}}{B_{y0}}$, plus some small deviation $\delta M$, or $M = M_{eq} + \delta M$. Although $\delta M$ is small, relative to $M_{eq}$, its derivative isn't necessarily small and is kept. We will validate these assumptions in the next section. Under an expansion around $\delta M$,
\begin{equation}
    \frac{M(\tilde \psi_1 \Omega) M'(\tilde \psi_1 \Omega)}{(1-M(\tilde \psi_1 \Omega)^2)} = \frac{M_{eq} M_{eq}'}{(1-M_{eq}^2)} + \frac{M_{eq} \delta M'}{(1-M_{eq}^2)}.
\end{equation}
To lowest order in $W$, $\frac{M_{eq}}{(1-M_{eq}^2)}\frac{dM_{eq}}{d\Omega} \approx -\mu \tilde \psi_1$, and
\begin{equation}\label{eq:Matched (Psi) Final psi}
    \begin{split}
    \Delta'\tilde{\psi}_1^{1/2} &= 4 S \frac{\partial \tilde{\psi}_1}{\partial t} A + 8(1+\mu)\tilde{\psi}_1 A + 8 \tilde \psi_1 P
\end{split}
\end{equation}
where $A$ and $P$ are the integrals
\begin{equation}\label{eq:A}
    A = \int^{1}_{\Omega_-}d\Omega\frac{\langle\cos(Y)\rangle^2_\Omega}{\langle1\rangle_\Omega},
\end{equation}
\begin{equation}\label{eq:P}
    P = \int^{1}_{\Omega_-} \frac{d\Omega}{\tilde \psi_1} \frac{M_{eq}(\tilde\psi_1\Omega) \delta M'(\tilde\psi_1\Omega)}{(1-M_{eq}^2(\tilde\psi_1\Omega))}\left( \langle\cos^2(Y)\rangle_\Omega-\frac{\langle\cos(Y)\rangle_\Omega^2}{\langle1\rangle_\Omega}\right).
\end{equation}
We have used Eq.~\eqref{Appendix eq:cos to z} to eliminate the term $-4\mu z_\epsilon$ in Eq.~\eqref{eq:Matched J, Mach By exp}. This result is the same as if we had expanded $M(\psi)$ at the same time as $B_y$, defined in Eq.~\eqref{Appendix eq:By exp}. The first term on the right-hand side of Eq.~\eqref{eq:Matched (Psi) Final psi} and the one on the left recover Rutherford's model, which describes the nonlinear evolution of the constant-psi tearing mode \cite{Rutherford}. In the limit $\Omega_- \rightarrow -\infty$, the integral $A$ has the asymptotic value of $0.41$, as shown in other works \cite{Rutherford,MiPo,Fitzpatrick}. The effect of the flow enters nonlinearly through the remaining two terms on the right of Eq.~\eqref{eq:Matched (Psi) Final psi}, both of which impact the saturation of the magnetic island. The second term on the right-hand side is a modification of the saturation term derived in \citet{MiPo}, and is a consequence of retaining contributions to the next order in $W$. The factor $(1+\mu)$ is new to this paper and measures the contribution of the Alfv\'enic flow to this term. Hereafter, we will refer to this as the `saturation term'. The third term on the right-hand side represents the nonlinear polarization current induced by the flow, which is the main result of this paper.  In principle, one could calculate an asymptotic value for $P$ if one knew the form of $M(\tilde{\psi}_1\Omega)$, or $\delta M$. Note that our model is invariant to the viscosity. The viscosity plays the implicit role of shaping $\delta M$, but this does not enter the model, as it assumes one knows $\delta M$, either by construction or by determination from simulation or experimental data.

We seek to determine the contribution of the flow to the magnetic island size, $W$. 
Replacing everywhere $\tilde{\psi}_1$ with the island width definition, Eq.~\eqref{eq:Island Size}, we find
\begin{equation}\label{eq:Island Width Change}
    AS\frac{dW}{dt} = \frac{\Delta'}{2} - (1+\mu)WA - W P.
\end{equation}
In the following section, we test the model by comparing it to the results from numerical simulations.

\section{Numerical Simulations}\label{sec:Numerical Simulations}

Here, we benchmark the model we derived in the previous section, Eqs.~\eqref{eq:Matched (Psi) Final psi}-\eqref{eq:Island Width Change}, with numerical simulations. To this goal we use a fully nonlinear initial value code that integrates the RMHD equations \eqref{eq:Induction Equation}-\eqref{eq:Equation of Motion}. The code employs a semi-implicit scheme for time advancement based on an explicit third-order Runge-Kutta scheme, but with the resistive and viscous diffusion terms treated implicitly. Periodic boundary conditions are imposed along the $y$-direction while fields are set to zero at $z=\pm L_z$. Spatial derivatives in the $z$-direction are computed using a sixth-order compact pseudo-spectral method, and derivatives in the $y$-direction are evaluated using Fast Fourier Transform. Across simulations, we choose $L_z=5$, $n_z = 1024$, and $n_y = 64$. The box length along $y$, $L_y$, is varied between simulations, and the methodology is described in the following paragraph. Although the model is insensitive to the viscosity, a small viscosity, $R > S$, is used for numerical stability. For the simulations presented in this paper, we have chosen $S = 10^{4}$ and $R = 2 \times 10^{4}$. Table \ref{tbl:Parameters} contains a summary of the parameters for this study.

An initial perturbation is injected into the flux function as a series of low-amplitude modes, with amplitude $A_i = 0.001$, 
\begin{equation}
    \delta\psi = \sum_{n=1,N}
    \frac{A_i}{N}\exp{(-\frac{z^2}{2})\cos(k_n y +\varphi_n)},
\end{equation}
with $\varphi_n$ being a randomly generated phase, and the wavenumber $k_n = \frac{2\pi n}{L_y}$. The integer $N$ is provided as input. To test the model, we seek to allow the growth of one magnetic island, which exhibits constant-psi behavior, $\Delta' W\ll 1$, in the nonlinear stage. In practice, we pick $L_y$ such that the $n=1$ mode, with wavenumber $k_1$, is the fastest growing mode in the box. To verify that the constant-psi condition is met a priori, we wrote an eigenvalue solver to determine $\Delta^\prime$ as a function of $k$. The solver integrates Eq.~\eqref{eq:Alfven Resonances} for $\xi$ assuming $\gamma = 0$, which is valid in region III, from the boundary, with boundary condition $\xi(z_{bc}) = \exp(-k|z_{bc}|)$, towards $z=0$. From Eqs.~\eqref{eq:Ideal Linear Induction} and~\eqref{eq:Transverse Displacement}, we replace the transverse displacement for the perturbed flux function with $\tilde{\psi}_1 = -B_{y0}\xi$ to calculate $\Delta^\prime$ by its definition, Eq.~\eqref{eq:Delta’}. This process was necessary as an analytic solution for $\Delta^\prime$ in the presence of flow does not exist. In addition to informing our choice of $L_y$, this provides a handy comparison with the $\Delta^\prime$ measured from the simulations, which is necessary for testing the nonlinear model. Our choices for $L_y$ and our predicted and numerically measured values for $\Delta^\prime$ are provided in Table \ref{tbl:Parameters}.

In the following, we first present a case study of a plasma with a relatively large Mach number ($M_0=0.8$), in which the equilibrium flow plays a significant role in its evolution. We then summarize results across a range of equilibrium flow parameters $M_0$ and $r$. For most of this section, we examine cases in region III of the parameter space, where only the tearing mode is unstable. Finally, we also consider cases in region II with wavenumbers $k$ that are stable to KHI while the tearing mode is unstable and Alfv\'en resonances are present.

\begin{table}
    \centering
    \begin{tabular}{|c|c|c|c|c|c|c|}
        \hline
        $r$ & $M_0$ & $L_y$ $(k_1)$ &$\Delta^\prime$(pred.) &$\Delta^\prime$(sim.) & $\epsilon$ (\%) & Region\\
        \hline
        \multirow{3}{*}{0.8} & 0.1 & \multirow{3}{*}{1.67 (0.6) } & 2.11 &  1.85 & 3 & \multirow{14}{*}{III}\\
        \cline{2-2} \cline{4-6}
        & 0.2 & & 2.03 & 1.78 & 4&\\
        \cline{2-2} \cline{4-6}
        & 0.3 & & 1.89 & 1.67 &  4&\\ 
         \cline{2-6} 
        & 0.6 & 2.5 (0.4) & 2.00 & 1.60 & 8 &\\
        \cline{1-6} 
        \multirow{5}{*}{1} & 0 & \multirow{5}{*}{1.67 (0.6) } & \multirow{5}{*}{2.11} & 1.87 & 3 &\\ \cline{2-2} \cline{5-6}
        & 0.2 &  & & 1.87 & 4&\\
        \cline{2-2} \cline{5-6}
        & 0.4 &  & & 1.86 & 4&\\
        \cline{2-2} \cline{5-6}
        & 0.6 &  & & 1.83 & 5&\\
        \cline{2-2} \cline{5-6}
        & 0.8 &  & & 1.75 & 4&\\
        \cline{1-6}
        \multirow{6}{*}{1.5} & 0.2 & \multirow{5}{*}{1.43 (0.7)} & 1.50 & 1.31 & 3&\\
        \cline{2-2} \cline{4-6}
        & 0.4 & & 1.63 & 1.43 & 3&\\
        \cline{2-2} \cline{4-6}
        & 0.6 & & 1.91 & 1.70 & 4&\\
        \cline{2-2} \cline{4-6}
        & 0.8 & & 2.53 & 2.27 & 2 &\\
        \cline{2-3} \cline{4-6}
        & 0.9 & 1.25 (0.8) & 2.28 & 2.05 & 6 &\\
        \cline{2-7}
        & 1.02 & 1.00 (1.0) & 1.82 & 1.66 & 6 & \multirow{2}{*}{II}\\
        \cline{2-6}
        & 1.1 & 0.90 (1.1) & 2.21 & 1.95 & 17 &\\
        \hline
    \end{tabular}
    \caption{Numerical parameters and some measured quantities from the numerical simulations of this study. Across simulations, $L_z = 5$, for a domain along $z$ = $[-L_z,L_z]$, with grid points $n_z = 1024$. Along $y$, we pick $n_y = 64$ grid points. The resistivity and viscosity are initialized as $\eta = 10^{-4}$ and $\nu = 5 \times10^{-5}$.}
    \label{tbl:Parameters}
\end{table}

\subsection{Case Study: \texorpdfstring{$M_0 = 0.8$, $r = 1.5$}{M0 = 0.8, r = 1.5}}\label{subsec:Case Study}

We present here the case with equilibrium shear amplitude $M_0 = 0.8$ and width $r = 1.5$, which lies in the parameter space region III outlined in the linear analysis. 

For the present case, the results of the linear solver motivated us to pick $L_y = 1.44$ $(k_1=0.7)$ for constant-psi magnetic island growth. We report a measured value of $\Delta' = 2.27$. This is smaller than the predicted value listed in Table \ref{tbl:Parameters}, and it is a trend we notice across our simulations that we ascribe to numerical effects. Indeed, the analytic determination of $\Delta^{\prime}$ is conducted by asymptotically matching the outer solution of the ideal MHD equations across the resonant surface and, therefore, information about the resonant surface does not enter the calculation. The numerical solution, on the other hand, is a self-consistent solution to the resistive RMHD equations which necessarily has a finite-width inner layer. For finite resolution and finite resistivity, there is no clear separation between inner and outer layer. For this reason, there is a systematic deviation of the numerically calculated $\Delta^{\prime}$ from theory.

For constant-psi ordering, we require the amplitude of the first harmonic, $\tilde{\psi}_1$, to be much larger than the amplitudes of higher harmonics. The linear stage of constant-psi tearing is characterized by exponential growth of the first harmonic of $\tilde \psi_1$. When the magnetic island size, $W = \sqrt{\tilde{\psi}_1}$, becomes comparable with the linear layer width, $\delta$, the plasma dynamics are nonlinear and $\tilde \psi_1$ grows algebraically. The constant-psi approximation breaks down when $\Delta'W \sim 1$. 

\begin{figure}
    \centering

    \begin{subfigure}{0.32\linewidth}
        \centering
        \includegraphics[width=\linewidth]{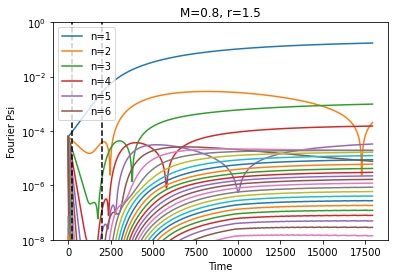}
        \caption{}
        \label{fig:Case Study-a}
    \end{subfigure}
    \hfill
    \begin{subfigure}{0.31\linewidth}
        \centering
        \includegraphics[width=\linewidth]{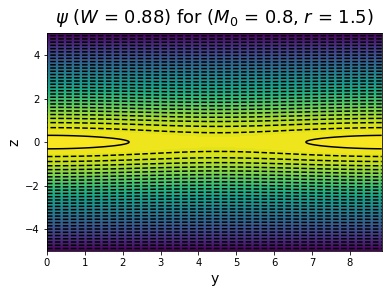}
        \caption{}
        \label{fig:Case Study-b}
    \end{subfigure}
    \hfill
    \begin{subfigure}{0.31\linewidth}
        \centering
        \includegraphics[width=\linewidth]{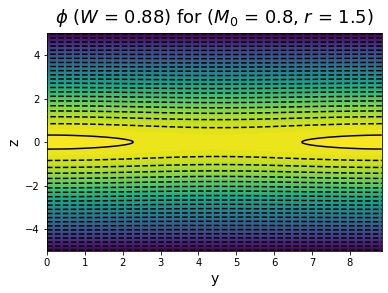}
        \caption{}
        \label{fig:Case Study-c}
    \end{subfigure}
    \caption{In panel (a), the simulated amplitudes of the harmonics $n=1-6$ of $\tilde{\psi}_n$ along the magnetic neutral line are plotted as a function of time. The vertical, dashed black lines mark the points between which we calculate the growth rate. Contours of the flux and stream functions, $\psi$ and $\phi$, during the nonlinear stage are presented in panels (b) and (c), respectively.}
    \label{fig:Case Study}
\end{figure}

The validity of the constant-psi ordering is verified in each simulation with a Fourier transform of the flux function along $y$ and a comparison of amplitudes. In Fig.~\ref{fig:Case Study-a}, we provide a log-linear plot of the amplitudes of the harmonics of the perturbed flux function. We observe that the first mode is dominant and grows exponentially during the linear stage. The growth is replaced with algebraic growth in the nonlinear stage, as expected. Contours of the flux function, $\psi$, and stream function, $\phi$, are plotted in Figs.~\ref{fig:Case Study-b} and~\ref{fig:Case Study-c}, respectively, when the magnetic island width has reached a value $W = 0.88$. The flux function has the expected structure for magnetic reconnection with an O-point at $y = 0$ and an X-point at $y = L_y/2$. The stream function has the same structure. This is consistent with Eq.~\eqref{eq:Phi Profile}, which predicted that, to lowest order, $\phi$ is constant along magnetic flux surfaces.

To benchmark our nonlinear model, we numerically evaluate the various terms appearing in Eq.~\eqref{eq:Island Width Change}. To this end, we interpolate the numerically simulated magnetic field $B_y$ and the spatial grid $z$ as functions of $\psi$ and $y$ in order to compute $A$ and $P$, defined in Eqs.~\eqref{eq:A} and~\eqref{eq:P}. From Eq.~\eqref{eq:Phi Profile}, we predict that the Mach number $M = \frac{d\Phi}{d\psi}$ is independent of $y$. We thus pick the Mach number across the magnetic island O-point, $M(y_o,z) = \frac{V_y(y_o,z)} {B_y(y_o,z)}$, for the numerical integration of $P$. The $\delta M^\prime$ term in the integrand of Eq.~\eqref{eq:P} is obtained by subtracting the known function $M_{eq}(\psi) = \frac{V_{y0}}{B_{y0}}$ from the interpolated $M(\psi)$ and then evaluating its derivative. The flux surface averages $\langle\cdot\rangle$ are carried out along constant flux surfaces in the region bounded by $W \ll z_{\epsilon} \ll 1$. Outside of the magnetic island, flux surface averages are carried out over the whole domain in $y$. The modification of magnetic field lines due to reconnection leads to magnetic flux loops that do not extend the full range of $y$, and the points at which they cross the magnetic neutral line constitute turning points of the magnetic field. These turning points are determined numerically, and flux surface averages are carried out between these turning points inside the magnetic island. The island width, $W(t)$, is evaluated from the first harmonic of the flux function, $\tilde{\psi}_1(t)$, using the definition in Eq.~\eqref{eq:Island Size}, while $\Delta^\prime$ is obtained from the jump in its derivative, as given in Eq.~\eqref{eq:Delta’}. The Simpson's method is used for numerical integration, and this process is repeated at each time step of the simulation.

\begin{figure}
    \centering

    \begin{subfigure}{0.33\linewidth}
        \centering
        \includegraphics[width=\linewidth]{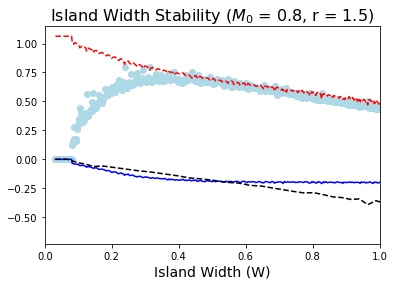}
        \caption{}
        \label{fig:Model M_0.8-a}
    \end{subfigure}
    \hfill
    \begin{subfigure}{0.32\linewidth}
        \centering
        \includegraphics[width=\linewidth]{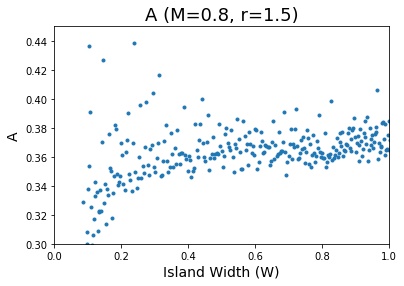}
        \caption{}
        \label{fig:Model M_0.8-b}
    \end{subfigure}
    \hfill
    \begin{subfigure}{0.32\linewidth}
        \centering
        \includegraphics[width=\linewidth]{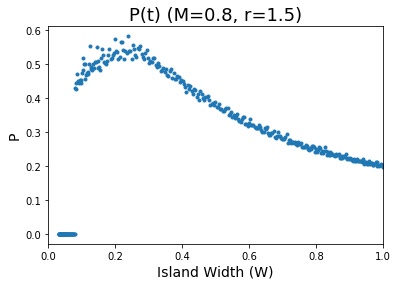}
        \caption{}
        \label{fig:Model M_0.8-c}
    \end{subfigure}
    \caption{In (a), we compare the terms of the magnetic island evolution equation, Eq.~\eqref{eq:Island Width Change}, for $M_0 = 0.8$ and $r = 1.5$. The lightly-toned blue dots represent the change in the island width on the left-hand side of the equation. The dashed red line is the contribution of each term on the right-hand side of the model. The contribution of the flow is depicted with the solid blue line. The dashed black line represents the saturation term, $-(1+\mu) WA$.  The evaluations of the integrals $A$ and $P$, Eqs.~\eqref{eq:A} and~\eqref{eq:P}, are provided, shown in (b) and (c).}
    \label{fig:Model M_0.8, r_1.5}
\end{figure}

Each term of the magnetic island evolution equation, Eq.~\eqref{eq:Island Width Change}, is compared in Fig.~\ref{fig:Model M_0.8-a}. The lightly-toned blue dots represent the left-hand side of Eq.~\eqref{eq:Island Width Change}. The right-hand side of the equation is represented by the dashed red line. The numerical agreement of the model is determined by the overlap of these two lines. The solid blue line shows the contribution of the polarization current, $-WP$, and the dashed black line represents the saturation term, $-(1+\mu) WA$.  As can be seen from Fig.~\ref{fig:Model M_0.8-a}, the polarization current (solid blue line) slows the magnetic island's growth during the nonlinear stage. The saturation term, $\propto-(1+\mu)$ with $\mu>0$, is also negative (black-dashed line) and it is larger in magnitude than in the case without flow, yielding an overall faster decrease of the nonlinear growth rate of the island. The model holds very well in the expected regime where $\delta < W$ and $\Delta' W < 1$, which parametrizes the nonlinear stage of the constant-psi tearing mode. To measure the accuracy of the model, we calculate the relative error, $\epsilon$, between the dashed red line and the light blue dots with the definition
\begin{equation}
    \epsilon = \frac{|AS\frac{dW}{dt}-\frac{\Delta'}{2} + (1+\mu)WA + WP|}{AS\frac{dW}{dt}}\times 100
\end{equation}
For this case, we find an $\epsilon$ = 2\%. We provide the relative error for all of our simulations in Table \ref{tbl:Parameters}.

The numerical values of the integrals $A(t)$ and $P(t)$ are provided in Figs.~\ref{fig:Model M_0.8-b} and~\ref{fig:Model M_0.8-c}, respectively. We note that the calculated value of $A$ is smaller than the expected asymptotic value of 0.411 reported in the literature. This discrepancy is caused by our numerical approach. In our analysis, we retain the entire field $B_y$ when evaluating flux integrals, rather than using the approximation in Eq.~\eqref{eq:By constant-psi}. The relatively large variation in $A$ is mainly due to the interpolation required to locate the turning points of magnetic field lines; this variation decreases with increasing resolution in $y$.

From the overlap of the dashed, red line and lightly-toned blue dots in Fig.~\ref{fig:Model M_0.8-a}, which have a mean relative error $\epsilon$ = 2\% for $W > 0.4$, we find that the model holds very well for this case.  To further validate our model, we have performed the analysis outlined here across a range of cases with different $M_0$ and $r$ values, and present the results in the following subsection.

\subsection{Model Validation Across Flow Parameters \texorpdfstring{$M_0$ and $r$}{M0 and r} Without Resonances}\label{subsec:Case Comp}

\begin{figure}[t]
    \centering

    \begin{subfigure}{0.28\linewidth}
        \centering
        \includegraphics[width=\linewidth]{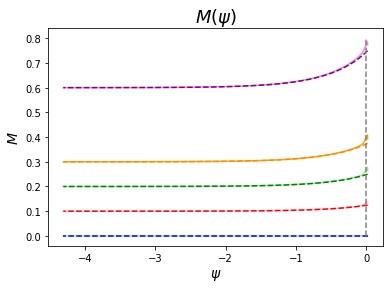}
        \caption{}
        \label{fig:M_psi-a}
    \end{subfigure}
    \hfill
    \begin{subfigure}{0.29\linewidth}
        \centering
        \includegraphics[width=\linewidth]{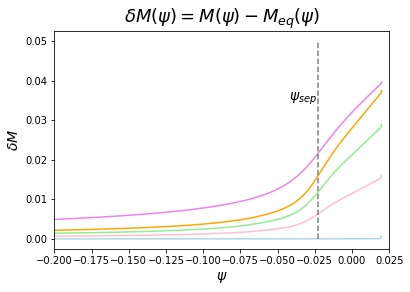}
        \caption{}
        \label{fig:M_psi-b}
    \end{subfigure}
    \hfill
    \begin{subfigure}{0.38\linewidth}
        \centering
        \includegraphics[width=\linewidth]{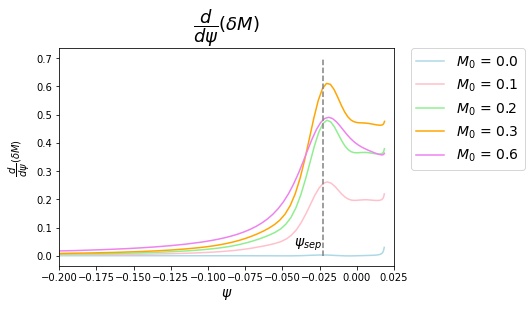}
        \caption{}
        \label{fig:M_psi-c}
    \end{subfigure}
    \caption{In (a), the simulated Mach number, $M(\psi) = \frac{V_y}{B_y}$, and reference state Mach number, $M_{eq}(\psi) = \frac{V_{y0}}{B_{y0}}$, are plotted with solid and dashed lines, respectively, for different values of $M_0$ at $r=0.8$. Their difference, $\delta M$, is plotted in (b). The derivative of $\delta M$ is plotted in (c). Each curve was plotted at $W = 0.6$.}
    \label{fig:M_psi}
\end{figure}

In this section, we test the model with the numerical results for the cases $M_0 = 0-0.6$ for $r=0.8$, and $M_0 = 0-0.8$ for $r=1.0$ and $r=1.5$ listed in Table~\ref{tbl:Parameters}. Our model assumes that the total (nonlinear) Mach number, $M = \frac{V_y}{B_y}$, deviates only slightly from the reference state Mach number, $M_{eq}$, such that $\delta M \ll M_{eq}$, where $\delta M$ measures this deviation. The derivatives of $\delta M$, however, are comparable to $M_{eq}$ and thus must be kept. These assumptions are supported by the simulation results shown in Fig.~\ref{fig:M_psi}, where we present data for $r=0.8$ with different values of $M_0$. In Fig.~\ref{fig:M_psi-a}, we plot $M$ and $M_{eq}$ as solid and dashed lines, respectively. Their difference, $\delta M$, and its derivative, $\delta M^\prime$ are shown in Fig.~\ref{fig:M_psi-b} and \ref{fig:M_psi-c}, respectively. As can be seen, the assumed ordering $\delta M\ll M_{eq}\sim M$, $\delta M^\prime\sim M$ is satisfied, supporting our model.

\begin{figure}
    \centering

    \begin{subfigure}{0.3\linewidth}
        \centering
        \includegraphics[width=\linewidth]{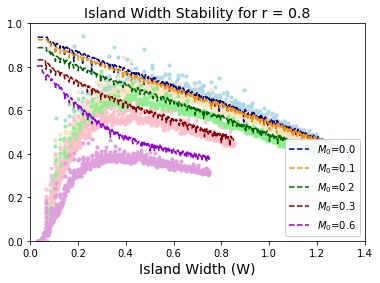}
        \caption{}
        \label{fig:Model r0.8-a}
    \end{subfigure}
    \hfill
    \begin{subfigure}{0.3\linewidth}
        \centering
        \includegraphics[width=\linewidth]{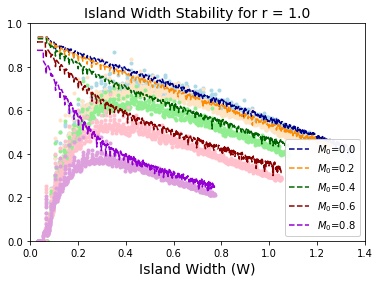}
        \caption{}
    \end{subfigure}
    \hfill
    \begin{subfigure}{0.3\linewidth}
        \centering
        \includegraphics[width=\linewidth]{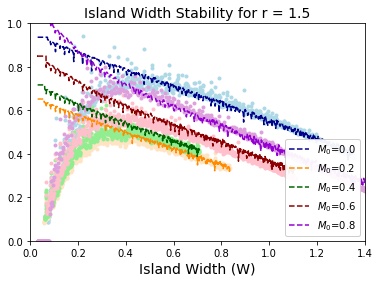}
        \caption{}\label{fig:Island Width Stability-c}
    \end{subfigure}
    \caption{The left-hand and right-hand sides of Eq.~\eqref{eq:Island Width Change}, which describes the rate of change of the magnetic island width $W$, are plotted here from left to right for $r= 0.8, 1.0$, and $r=1.5$. The lightly toned dots show the left-hand side of Eq.~\eqref{eq:Island Width Change}, $AS\frac{dW}{dt}$. The dark, dashed lines that overlap the dots at larger island widths represent the sum of the terms on the right-hand side of Eq.~\eqref {eq:Island Width Change}, where each shade of color indicates a different value of $M_0$.}
    \label{fig:Island Width Stability}
\end{figure}

\begin{figure}
    \centering

    \begin{subfigure}{0.3\linewidth}
        \centering
        \includegraphics[width=\linewidth]{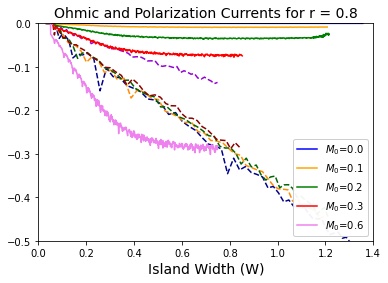}
        \caption{}
    \end{subfigure}
    \hfill
    \begin{subfigure}{0.3\linewidth}
        \centering
        \includegraphics[width=\linewidth]{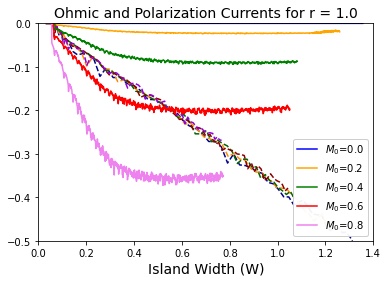}
        \caption{}
    \end{subfigure}
    \hfill
    \begin{subfigure}{0.3\linewidth}
        \centering
        \includegraphics[width=\linewidth]{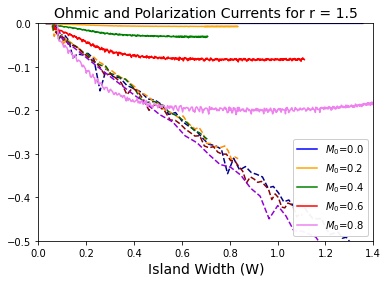}
        \caption{}
    \end{subfigure}
    \caption{The second term on the right-hand side of Eq.~\eqref{eq:Island Width Change}, $-(1+\mu)WA$, is plotted here with dashed lines for $r= 0.8, 1.0$, and $r=1.5$ from left to right. The solid lines show the simulated polarization current, which is represented by the third term on the right hand-side of Eq.~\eqref{eq:Island Width Change}, $-WP$.}
    \label{fig:Flow Currents}
\end{figure}

The results of the numerical simulation are summarized in Figs.~\ref{fig:Island Width Stability} and~\ref{fig:Flow Currents}. In Fig.~\ref{fig:Island Width Stability} we plot $AS\frac{dW}{dt}$ at each instant with the lightly toned dots. The different colors indicate the flow amplitude $M_0$. The darker, dashed lines show the sum of the terms on the right-hand side of the model, Eq.~\eqref{eq:Island Width Change}. In Fig.~\ref{fig:Flow Currents}, the solid lines represent the effect of the polarization current $-WP$ and the dashed lines represent the saturation term, $-(1+\mu)WA$.  From these plots, we note a few results. First, the lightly toned dots, which correspond to the left-hand side of Eq.~\eqref{eq:Island Width Change}, overlap with the dashed lines, corresponding to the right-hand side, with very good agreement, as indicated by $\epsilon$ in Table ~\ref{tbl:Parameters}, providing strong support to our model.  Second, the net contribution of the flow during the nonlinear stage, which is measured by the terms containing $\mu$ and $P$, is to slow the magnetic island's growth.

In \citet{MiPo}, the authors derived the term that quantifies the effect of saturation in the absence of an equilibrium flow  that, in this context, is $-WA$. Here we have derived the contribution of the Alfv\'enic flow to that term, which takes the form of an additive factor $-\mu AW$ for a total contribution of $-(1+\mu)WA$.  Relative to \citet{MiPo}, the effect of the flow is as follows. For $r=1$, the term remains unchanged from the no-flow case. Therefore, as for the polarization current term, it acts to slow the growth of the magnetic island. For $r=1.5$ ($r>1$), the factor $(1+\mu)$ is always greater than unity and increases with $M_0$. Consequently, this term always acts to slow the island growth, and its stabilizing effect becomes stronger as $M_0$ increases. For $r=0.8$ ($r<1$), $\mu<0$ and the factor $(1+\mu)$ decreases with increasing $M_0$, becoming negative for \begin{equation} M_0>\frac{\sqrt{3}r^2}{\sqrt{2+r^2}}. \end{equation} Therefore, for $r<1$, the stabilizing effect of the saturation term weakens as $M_0$ increases and, for sufficiently large $M_0$, the term may become destabilizing and act to increase the island growth rate. However, for the parameter range considered in our simulations, the saturation term is always stabilizing. 

\begin{figure}
    \centering

    \begin{subfigure}{0.27\linewidth}
        \centering
        \includegraphics[width=\linewidth]{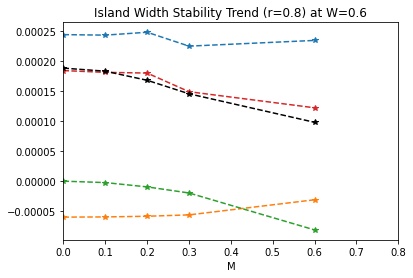}
        \caption{}
    \end{subfigure}
    \hfill
    \begin{subfigure}{0.27\linewidth}
        \centering
        \includegraphics[width=\linewidth]{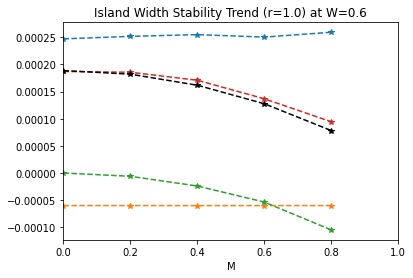}
        \caption{}
        \label{fig:Island Width Scales-b}
    \end{subfigure}
    \hfill
    \begin{subfigure}{0.39\linewidth}
        \centering
        \includegraphics[width=\linewidth]{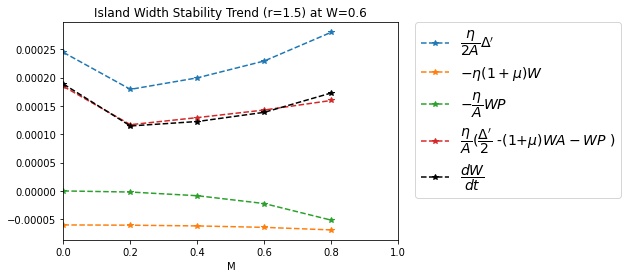}
        \caption{}
    \end{subfigure}
    \caption{Plots of the various terms entering Eq.~\eqref{eq:Island Width Change} as a function of $M_0$, taken when the island size has reached $W = 0.6$, are presented here for $r=$ 0.8, 1.0, 1.5. The left and right-hand sides of Eq.~\eqref{eq:Island Width Change} are shown with the black and red lines, respectively. The light blue line shows the contribution of $\Delta^\prime$, the contribution of the polarization current $P$ is shown in green, and the saturation term $-(1+\mu)WA$ with yellow.}
    \label{fig:Island Width Scales}
\end{figure}

To provide another perspective on the contribution of the various terms in the model, in Fig.~\ref{fig:Island Width Scales} we plot a cut of the terms shown in Figs.~\ref{fig:Island Width Stability} and~\ref{fig:Flow Currents} for fixed $W = 0.6$. From the left panel to the right, we plot these cuts for $r=0.8$, 1.0, and 1.5. The dashed blue line visualizes the contribution of $\Delta^\prime$. The green and yellow lines show the contributions of the polarization current and the saturation term, respectively. The dashed black line shows the left-hand side measuring $AS\frac{d W}{dt}$ and the dashed red line is the sum of terms on the right-hand side. We observe the same trends as before: the polarization current is increasingly negative with increasing flow amplitude $M_0$, and the net nonlinear effect of the flow is to slow-down the growth of the magnetic island. We observe that the island growth rate increases with $M_0$ for $r>1$ due to the dominant contribution of the flow-modified $\Delta^\prime$ term.

There is a disagreement for the case $M=0.6$ and $r=0.8$ in Fig.~\ref{fig:Model r0.8-a} with an error $\epsilon=8\%$ reported in Table \ref{tbl:Parameters}. The mismatch is a consequence of dropping $\delta M$ relative to $M_{eq}$ in Eq.~\eqref{eq:P}. By retaining that factor, we find $\epsilon$ = 5\%. Performing the same operation improves all simulated cases. However, this is a small correction, and Eq.~\eqref{eq:P} is sufficient for determining the impact of the equilibrium flow on the magnetic island's growth. 

There is another source of disagreement in the polarization integral \eqref{eq:P} due to $M_{eq}$. Strictly speaking, the factor $\frac{M_{eq} \delta M'}{(1-M_{eq}^2)}$ is not expanded to lowest order in $W$ here, as $\delta M$ is defined by $M = M_{eq} + \delta M$ and the analytic size of $\delta M^\prime$ remains unclear. To leading order, simulations indicate that $\delta M$ is of order one, implying that the polarization term is of the same order as $\Delta^\prime$. Yet, terms of order $O(W^2)$ remain in the numerical solution of $\delta M^\prime$, but under constant-psi ordering these terms are small and we do not believe they are a source of significant error in the matching of the numerical results. Finally, in the analysis of simulations, we use the entire field $B_y$ in the flux surface integration, \eqref{eq:Flux Surface Average}, as opposed to \eqref{eq:By constant-psi}. This made integration near the separatrix numerically tractable, and we anticipated the impact of retaining higher orders to be small.

\subsection{Alfv\'en Resonances}\label{subsec:Resonances}

So far, our analysis has been concerned with cases in parameter space region III, in which the flow is everywhere sub-Alfv\'enic, and the tearing mode is the only energy-converting process present. The tearing mode is also allowed to be present in region II, in which the equilibrium flow profile is sub-Alfv\'enic around the magnetic neutral line, but is asymptotically super-Alfv\'enic. Alfv\'en resonances form as one moves from region III to region II and extract energy that would otherwise have gone to the tearing mode. This behavior differs from that found by \citet{Li2016} in that their magnetic islands were formed due to nonlinear KHI, and are not describable by constant-psi tearing mode theory. As we are concerned with magnetic islands that form due to the resistive tearing mode, we strategically select wavenumbers of the first harmonic, $k_1$, for which the KHI is stable, and the tearing mode parameter is of order unity for constant-psi behavior. In this section, we look at two new cases which exhibit resonances: $M_0 = 1.02$ and $M_0 = 1.1$ for $r=1.5$. The cases $M_0 = 0.9$ and $M_0 = 0.8$ for $r=1.5$ and $M_0 = 0$ for $r=1$ are plotted also for comparison. The initial conditions and important parameters are provided in Table \ref{tbl:Parameters}.

\begin{figure}
    \centering

    \begin{subfigure}{0.32\linewidth}
        \centering
        \includegraphics[width=\linewidth]{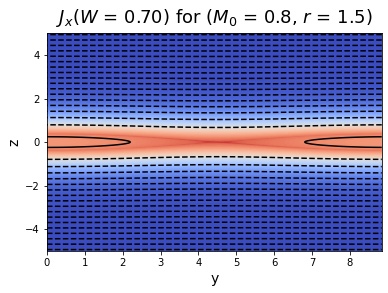}
        \caption{}
        \label{fig:J M0.8}
    \end{subfigure}
    \hfill
    \begin{subfigure}{0.32\linewidth}
        \centering
        \includegraphics[width=\linewidth]{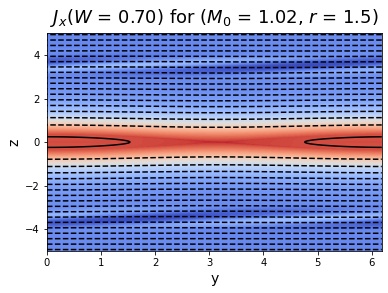}
        \caption{}
        \label{fig:J M1.02}
    \end{subfigure}
    \hfill
    \begin{subfigure}{0.32\linewidth}
        \centering
        \includegraphics[width=\linewidth]{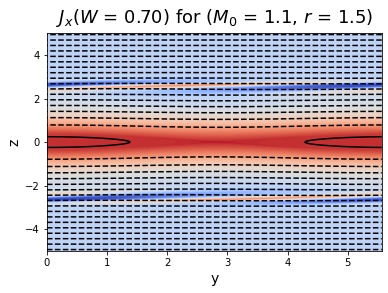}
        \caption{}
        \label{fig:J M1.1}
    \end{subfigure}
    \caption{Here, we present contours of the current densities for the cases with $r = 1.5$, $M_0 = $ (a) $0.8$, (b) $1,02$, and (c) $1.1$. In (b) and (c), resonances form where the flow transitions from sub-Alfv\'enic to super-Alfv\'enic flow.}
    \label{fig:J Resonances}
\end{figure}

Contours of the current density in the nonlinear stage for the cases $M_0 = 0.8$, 1.02, and 1.1 for $r=1.5$ are presented in Fig.~\ref{fig:J Resonances}. We observe that the resonances form during the linear stage at the expected locations, which are determined by the resonance conditions $V_{y0}(z) = \pm V_A(z)$.

\begin{figure}
    \centering

    \begin{subfigure}{0.36\linewidth}
        \centering
        \includegraphics[width=\linewidth]{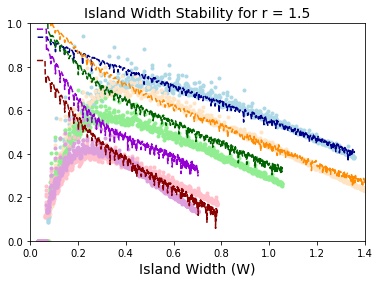}
        \caption{}
    \end{subfigure}
    \begin{subfigure}{0.45\linewidth}
        \centering
        \includegraphics[width=\linewidth]{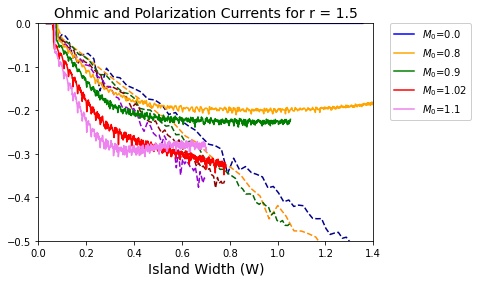}
        \caption{}
    \end{subfigure}
    \caption{(a) The left-hand and right-hand sides of Eq.~\eqref{eq:Island Width Change}, which describes the rate of change in the magnetic island width, are plotted here for $r=1.5$. Alfv\'en resonances are excited for the cases $M_0=1.02$ and $M_0=1.1$. The lightly toned dots show the left-hand side of the equation, $AS\frac{dW}{dt}$. The dark, dashed lines that overlap the dots at larger island widths are the sums of terms on the right side of the equation. (b) The dashed lines show the saturation term, which is represented by the second term on the right-hand side of Eq.~\eqref{eq:Island Width Change}, $-(1+\mu)WA$. The effects of the polarization current, which is represented by the third term on the right-hand side of Eq.~\eqref{eq:Island Width Change}, $-WP$, are shown with solid lines.}
    \label{fig:Matching Resonances}
\end{figure}

We apply the model, Eq.~\eqref{eq:Island Width Change}, to cases which excite Alfv\'en resonances ($M_0 >1$) and compare them to cases which do not ($M_0 < 1$). The results are summarized in Fig.~\ref{fig:Matching Resonances}. We find that the model holds for the weakly resonant case, $M_0 = 1.02$, for which the resonances develop far from the tearing layer, but breaks down for $M_0 = 1.1$, for which the resonances are comparatively closer to the magnetic island. Since the resonances develop during the linear stage of the magnetic island's evolution, they are well established when the island transitions to the nonlinear stage. Although we do not investigate here how resonances affect the nonlinear growth of tearing mode islands, these resonances act as a barrier to the magnetic island's growth. With increasing $M_0$, we anticipate the model to break down further, as the island is no longer described by constant-psi.

\section{Conclusions}\label{subsec:Conclusions}

We have explored the linear and nonlinear evolution of a plasma governed by the reduced Magnetohydrodynamics model, Eqs.~\eqref{eq:Induction Equation}-\eqref{eq:Equation of Motion}, in the presence of both a sheared equilibrium magnetic field and flow, Eqs.~\eqref{eq:B0}-\eqref{eq:V0}. In the linear analysis, we identified four regions, based on the nature and occurrence of Alfv\'en singularities, in the parameter space spanned by the equilibrium Alfv\'enic flow amplitude, $M_0$, the ratio of the velocity to magnetic shear width, $r$, and the mode wave number, $k$, to understand under which conditions the KHI, TM, and Alfv\'en resonances are present.
Four regions are defined:

\begin{itemize}
    \item \textbf{Region I}  \textbf{($\mathbf{1 < M_0}$, $ \mathbf{r < M_0}$)}: the flow is everywhere super-Alfv\'enic and which the KHI are allowed.
    
    \item \textbf{Region II} \textbf{( $\mathbf{ 1 < M_0 < r}$)}: the flow is asymptotically super-Alfv\'enic and resonances form at the boundary separating the super and sub-Alfv\'enic regions. KHI and TM are allowed.
    
    \item \textbf{Region III} \textbf{($\mathbf{M_0 < 1}$, $ \mathbf{M_0 < r}$)}: the flow is everywhere sub-Alfv\'enic and TM is allowed.
    
    \item \textbf{Region IV} \textbf{( $\mathbf{ r < M_0 < 1}$)}: the flow is asymptotically sub-Alfv\'enic with a super-Alfv\'enic region lying between the Alfv\'en resonances. KHI is allowed.
\end{itemize}
In region III, which is spanned by $M_0 < 1$ and $M_0 < r$, the tearing mode is the only process that can be induced. This made region III a great candidate for studying magnetic island formation.

We extended Rutherford's model, which describes the nonlinear evolution of constant-psi magnetic islands, to describe the evolution of magnetic islands in the presence of a vortex sheet. The main result of this work is Eq.~\eqref{eq:Island Width Change}. We verified with nonlinear simulations that the model held in region III, and even for some cases in region II, in which Alfv\'en resonances are present. 

We found that the flow enters the model nonlinearly in two ways: through a polarization current $P$ and a term $\mu$ that modifies the saturation term derived in \citet{MiPo}. It is found that $P$ slows the magnetic island's growth. Meanwhile, the effect of flow on the saturation term, $\propto-(1+\mu)$, depends on $r$. For $r=1$, it remains unmodified by the flow. For $r > 1$, its magnitude increases with $M_0$, an effect which further slows the magnetic island growth. For $r < 1$, the magnitude of this term decreases with $M_0$ and the factor $-(1+\mu)$ becomes positive for $M_0>\sqrt{3}r^2/(\sqrt{2+r^2})$, transitioning from stabilizing to destabilizing the island's growth. However, we find that the total nonlinear contribution of the flow is stabilizing in the considered parameter range.

In addition to affecting the model through \(P\) and \(\mu\), the flow also modifies the linear tearing mode stability parameter $\Delta^{\prime}$, which enters the algebraic growth term of the Rutherford equation, when \(r \neq 1\). For \(r < 1\), \(\Delta^{\prime}\) decreases with increasing \(M_0\), leading to a reduction in the island growth rate. In contrast, for \(r > 1\), \(\Delta^{\prime}\) increases with \(M_0\), thereby enhancing the nonlinear growth of the island and competing with the stabilizing effects of the polarization current and saturation terms. As can be seen from Fig. \ref{fig:Island Width Scales}, the relative sizes of the latter terms with respect to the $\Delta^{\prime}$ term depend on $M_0$ and $r$ but generally reach the same order of magnitude as $M_0$ increases. For the case $M_0=0,8$ and $r=1$, the polarization current and saturation term are about $40\%$ and $20\%$ of the $\Delta^{\prime}$ term at $W=0.6$, respectively. Meanwhile, for $M_0=0.8$ and $r=1.5$, the terms are comparatively smaller, standing at about $16\%$ and $20\%$ at $W=0.6$, respectively. This explains why in this case the island growth rate displays the same dependence on $M_0$ as $\Delta^\prime$.

In the parameter space region II, for which $1<M_0<r$, Alfv\'en resonances are present, and the effect of increasing the Alfv\'enic Mach number is to generate the resonances ever closer to the magnetic neutral line. As the resonances approach the neutral line, they increase in magnitude and act as magnetic barriers inhibiting the magnetic island's growth. The coupling of the magnetic island with the resonances due to magnetic forces leads to a modification of the nonlinear island growth, which is no longer adequately described by constant-psi. As a consequence, this effect is not captured in our model. We observe this deviation with our numerical simulations, with the results summarized in Fig.~\ref{fig:Matching Resonances}.  We defer a detailed investigation of the interplay between Alfv\'en resonances  and nonlinear magnetic island dynamics to future work.

This work was conducted with relatively large resistivity and viscosity, $\eta = 10^{-4}$, and $\nu = 5 *10^{-5}$. We anticipate better agreement between the simulations and the model for smaller $\eta$ and minimized $\nu$. Additionally, higher resolution along $y$ would reduce the variability in the numerical interpolation of the flux surface averages, defined by Eq.~\eqref{eq:Flux Surface Average}, and lead to more consistent evaluations of $A$ and $P$, or Eqs.~\eqref{eq:A} and~\eqref{eq:P}.

The model equations, Eqs.~\eqref{eq:Matched (Psi) Final psi}-\eqref{eq:Island Width Change}, which retain terms up to order $W$, provide an approximation for the evolution of constant-psi magnetic islands in the presence of an Alfv\'enic flow. One limitation of our model, however, is that we don't have an analytic expression for $\delta M$, the deviation of the Mach number from the reference state. Although $\delta M$ is small relative to the reference state Mach number, $M_{eq}$, our numerical results showed that its derivative, $\delta M ^\prime$, was the same order in $W$ as $M_{eq}^\prime$ and provided a sizable contribution to the magnetic island's growth. Having an analytic solution for $\delta M$ would allow one to write an analytic expression for the polarization integral $P$, appropriate to the lowest order in $W$. Then, one could determine the magnitude and shape of $P$ and have a complete analytic expression for the magnetic island's width. This is useful for both experimentalists and theorists. For example, an experimentalist measuring magnetic islands could determine the applicable model from their growth, and for constant-psi magnetic islands, the experimentalist would have the shape of $\delta M$ without a direct measurement. A theorist running simulations would also be able to focus their efforts on the physics of interest and have an analytic representation of the flow available without the need to simulate it. Given the relevance of Alfv\'enic flows in space and laboratory environments, an important direction for future work is to find an analytic expression for $\delta M$.

\begin{acknowledgments}
This paper was supported by NSF grant 2108320 and by the US DOE under grant No. DE-FG02-04ER54742. One of us (AT) was also supported by NSF CAREER grant 2141564. We also acknowledge the Texas Advanced Computing Center (TACC) at The University of Texas at Austin for providing HPC resources that have contributed to the research results reported within this paper. 
\end{acknowledgments}

\appendix
\section{}
Here we provide the intermediate steps for the derivation of Eq.~\eqref{eq:Matched (Psi) Final psi}, the modified Rutherford model which accounts for the presence of flow.

Substituting in our asymptotic solution to $\psi$ in the outer region, Eq.~\eqref{eq:Psi Outer}, and the inner layer solution for $J$, Eq.~\eqref{eq:J(M)}, into the matching condition \eqref{eq:Matching} provides,
\begin{equation}
\begin{split}
    \Delta' - 4(1+\mu)z_{\epsilon}  &= -\frac{1}{\pi \tilde{\psi}_1} \oint dY \int^{\psi_+}_{\psi_-} \frac{d\psi}{B_y} \left(\frac{S\langle\frac{\partial\psi}{\partial t}\rangle}{\langle1\rangle} + \frac{\langle J_0 \rangle}{\langle 1 \rangle}\right)\cos(Y)
    \\&-\frac{1}{\pi \tilde{\psi}_1} \oint dY \int^{\psi_+}_{\psi_-}\frac{d\psi}{B_y}\frac{M(\psi) M'(\psi)}{(1-M^2(\psi))}\left(B_{\perp}^2-\frac{\langle B_{\perp}^2\rangle}{\langle1\rangle} \right)\cos(Y) ,
\end{split}
\end{equation}
where $\psi_{\pm} = \psi(\pm z_{\epsilon})$. In addition to substitution, we have changed variables from $z$ to $\psi$ with the Jacobian $dz = \frac{dz}{d\psi}d\psi = -\frac{d\psi}{B_y}$ since Eq.~\eqref{eq:J(M)} specifies $J$ in terms of the variables $\psi$ and $y$ (or $Y$). The $k^2$ dependent term in the outer solution, $\tilde{\psi}_{out}$, has been canceled with the term corresponding to $\frac{d^2\tilde\psi_{out}}{dy^2}$ due to the periodic nature of the amplitude, defined with Eq.~\eqref{eq:psi out Amp}. Replacing integrals in $Y$ with the flux surface average operator defined in \eqref{eq:Flux Surface Average}, we obtain the expression
\begin{equation}
\begin{split}
   \Delta' - 4(1+\mu)z_{\epsilon} &= -\frac{2}{ \tilde{\psi}_1}  \int^{\psi_+}_{\psi_-}d\psi  \langle\left(\frac{S\langle\frac{\partial\psi}{\partial t}\rangle}{\langle1\rangle} + \frac{\langle J_0 \rangle}{\langle 1 \rangle} \right)\cos(Y) \rangle
    \\& - \frac{2}{\tilde{\psi}_1} \int^{\psi_+}_{\psi_-}d\psi\frac{M(\psi) M'(\psi)}{(1-M^2(\psi))}\langle\left(B_{\perp}^2-\frac{\langle B_{\perp}^2\rangle}{\langle1\rangle} \right)\cos(Y)\rangle ,
\end{split}
\end{equation}
The first term on the right here is reminiscent of equation 14 in \citet{Rutherford}. We recover Rutherford's integral with the constant-psi approximation, Eq.~\eqref{eq:Psi Constant-psi}, and a reminder that the reference fields are sustained by an external source such that $\frac{d\psi_{0}}{dt} = 0$,
\begin{equation}\label{Appendix eq:Matched J0}
    \begin{split}
    \Delta' - 4(1+\mu)z_{\epsilon} &= -\frac{2S}{ \tilde{\psi}_1} \frac{d\tilde{\psi}_1}{dt}\int^{\psi_+}_{\psi_-}d\psi\frac{\langle\cos(Y)\rangle^2}{\langle1\rangle} -\frac{2}{ \tilde{\psi}_1} \int^{\psi_+}_{\psi_-}d\psi\left< \frac{\langle J_0\rangle}{\langle1\rangle} \cos(Y)\right>
    \\& - \frac{2}{\tilde{\psi}_1} \int^{\psi_+}_{\psi_-}d\psi\frac{M(\psi) M'(\psi)}{(1-M^2(\psi))}\left<\left(B_{\perp}^2-\frac{\langle B_{\perp}^2\rangle}{\langle1\rangle} \right)\cos(Y)\right>.
\end{split}
\end{equation}
Under constant-psi ordering, 
\begin{equation}\label{Appendix eq:By constant-psi}
    B_y = -\frac{d\psi}{dz}\approx \sign(z)\sqrt{-2(\psi - \tilde{\psi}_1\cos(Y))},
\end{equation}
And flux surface averages are appropriately
\[ \begin{cases}
    \begin{aligned}
      \langle f\rangle = \frac{1}{\sqrt{\tilde \psi_1}}\langle f\rangle_\Omega &= \frac{1}{2\pi \sign(z)} \frac{1}{\sqrt{2 \tilde \psi_1 }}\int^{2\pi}_{0} \frac{f}{\sqrt{-\Omega + \cos(Y)}} dY & \Omega = \psi/\tilde\psi_{1} > 1,\\
      &=\frac{1}{2\pi \sign(z)} \frac{1}{\sqrt{2 \tilde \psi_1 }}\int^{2\pi-Y_{tp}}_{Y_{tp}} \frac{f}{\sqrt{-\Omega + \cos(Y)}} dY & \Omega < 1,\\
      \end{aligned}
   \end{cases}
\]
where $Y_{tp} = \arccos{\Omega}$ with the normalized flux function defined as $\Omega = \psi/\tilde\psi_1$. With this definition, the model equation is written as
\begin{equation}\label{Appendix eq:Matched J0 Normalized}
    \begin{split}
    \Delta' - 4(1+\mu)z_{\epsilon} &= -\frac{2S}{ \tilde{\psi}_1^{1/2}} \frac{d\tilde{\psi}_1}{dt}\int^{\Omega_+}_{\Omega_-}d\Omega\frac{\langle\cos(Y)\rangle^2_\Omega}{\langle1\rangle_\Omega}
    -\frac{2}{ \tilde{\psi}_1^{1/2}} \int^{\Omega_+}_{\Omega_-}d\Omega\left< \frac{\langle J_0\rangle_\Omega}{\langle1\rangle_\Omega}\cos(Y)\right>_\Omega
    \\& - \frac{2}{ \tilde{\psi}_1^{3/2}}\int^{\Omega_+}_{\Omega_-}d\Omega\frac{M(\tilde\psi_1\Omega) M'(\tilde\psi_1\Omega)}{(1-M^2(\tilde\psi_1\Omega))}\left<\left(B_{\perp}^2-\frac{\langle B_{\perp}^2\rangle_\Omega}{\langle1\rangle_\Omega} \right)\cos(Y)\right>_\Omega,
\end{split}
\end{equation}
and the derivative in $M'(\tilde{\psi}_1 \Omega)$ is now defined as a derivative with respect to $\Omega$. For the equilibrium profiles imposed, Eqs.~\eqref{eq:B0} and~\eqref{eq:V0}, $\Omega_{\pm}$ is negative. We assume symmetry across the neutral line and flip the integral in $\Omega$ to integrate from a negative boundary condition to the positive boundary condition provided at $z=0$, 
\begin{equation}\label{Appendix eq:Matched J0 Symmetric}
    \begin{split}
    \Delta' - 4(1+\mu)z_{\epsilon} &= \frac{4S}{ \tilde{\psi}_1^{1/2}} \frac{d\tilde{\psi}_1}{dt} A
    +\frac{4}{ \tilde{\psi}_1^{1/2}} \int^{1}_{\Omega_-}d\Omega\left< \frac{\langle J_0\rangle_\Omega}{\langle1\rangle_\Omega} \cos(Y)\right>_\Omega
    \\& + \frac{4}{ \tilde{\psi}_1^{3/2}} \int^{1}_{\Omega_-}d\Omega\frac{M(\tilde\psi_1\Omega) M'(\tilde\psi_1\Omega)}{(1-M^2(\tilde\psi_1\Omega))}\left<\left(B_{\perp}^2-\frac{\langle B_{\perp}^2\rangle_\Omega}{\langle1\rangle_\Omega} \right)\cos(Y)\right>_\Omega,
\end{split}
\end{equation}
and the factor $\sign(z)$ is positive in Eq.~\eqref{Appendix eq:By constant-psi}. $A$ here is Rutherford's integral
\begin{equation} \label{Appendix eq: Rutherford A}
    A = \int^{1}_{\Omega_-}d\Omega\frac{\langle\cos(Y)\rangle^2_\Omega}{\langle1\rangle_\Omega}.
\end{equation}

The second term on the right-hand side leads to the same term as derived in \citet{MiPo}, which in the absence of an Alfv\'enic flow determines the saturation of the island, and we follow their lead in deriving it here. The equilibrium current taken together with the constant-psi ordering, Eq.~\eqref{eq:Psi Constant-psi Expansion}, is, correct to order $O(W^2)$, 
\begin{equation}\label{Appendix eq:J0 Equilibrium}
    J_0 = -1 + z^2 + O(z^4) = -1 - 2\Omega \tilde \psi_1 + 2 \tilde \psi_1 \cos(Y)+O(\tilde \psi_1^2).
\end{equation}
Expansion of the second term in \eqref{Appendix eq:Matched J0 Symmetric} leads to
\begin{equation}
\begin{split}
    \left<\frac{\langle J_0\rangle_\Omega}{\langle1\rangle_\Omega} \cos(Y) \right>_\Omega &\approx \left<\cos(Y) + 2\Omega\tilde\psi_1 \cos(Y) -2\tilde \psi_1\frac{\langle\cos(Y)\rangle_\Omega}{\langle1\rangle_\Omega}\cos(Y) \right>_\Omega \\&= - \langle J_0 \cos(Y) \rangle + 2\tilde{\psi}_1\left( \langle\cos^2(Y)\rangle_\Omega-\frac{\langle\cos(Y)\rangle_\Omega^2}{\langle1\rangle_\Omega}\right).
\end{split}
\end{equation}
Integration of the flux surface average $J_0 \cos(Y)$ is zero due to the projection along the $n=1$ harmonic with $\cos(Y)$. Substitution of the remaining term into Eq.~\eqref{Appendix eq:Matched J0 Symmetric} then leads to the expression
\begin{equation}\label{Appendix eq:Matched J, J_0 exp}
    \begin{split}
    \Delta' - 4(1+\mu)z_{\epsilon} &= \frac{4S}{ \tilde{\psi}_1^{1/2}} \frac{d\tilde{\psi}_1}{dt} A 
    -8 \tilde{\psi}_1^{1/2} \int^{1}_{\Omega_-}d\Omega \langle\cos^2(Y)\rangle_\Omega
    +8 \tilde{\psi}_1^{1/2} A
    \\& + \frac{4}{ \tilde{\psi}_1^{3/2}} \int^{1}_{\Omega_-}d\Omega\frac{M(\tilde\psi_1\Omega) M'(\tilde\psi_1\Omega)}{(1-M^2(\tilde\psi_1\Omega))}\left<\left(B_{\perp}^2-\frac{\langle B_{\perp}^2\rangle_\Omega}{\langle1\rangle_\Omega} \right)\cos(Y)\right>_\Omega,
\end{split}
\end{equation}
It can be shown that
\begin{equation}\label{Appendix eq:cos to z}
\int^{1}_{\Omega_-}d\Omega\langle\cos^2(Y)\rangle_{\Omega} = \frac{1}{2\tilde{\psi}_1^{1/2}}z_{\epsilon},
\end{equation}
and we observe a cancellation of the second term on the left-hand side of Eq.~\eqref{Appendix eq:Matched J, J_0 exp}, $-4z_{\epsilon}$, with the second term on the right, $-8 \tilde{\psi}_1^{1/2} \int^{1}_{\Omega_-}d\Omega \langle\cos^2(Y)\rangle_\Omega$.

This model describes the evolution of the first harmonic of the magnetic flux, $\tilde{\psi}_1$, which is consistent with past literature and contains the contribution of the Alfv\'enic flow in the profile function $M(\psi)$. We will now simplify the last expression on the right, which measures the effect of the Alfv\'enic flow. By definition, $B_{\perp}^2 = B_y^2 + B_z^2$. In the inner layer, $\partial_z \gg \partial_y$ and the contributions from $B_z = \frac{d\psi}{dy}$ are negligible. Under constant-psi ordering, Eq.~\eqref{Appendix eq:By constant-psi}, the function under the flux surface average operator in the last integral of Eq.~\eqref{Appendix eq:Matched J, J_0 exp} is approximately
\begin{equation}
\begin{split}\label{Appendix eq:By exp}
    \langle\left(B_{y}^2-\frac{\langle B_{y}^2\rangle_\Omega}{\langle1\rangle_\Omega} \right)\cos(Y)\rangle_\Omega &\approx -2\left(\Omega \tilde\psi_1\langle \cos(Y)\rangle_\Omega-\tilde{\psi}_1 \langle\cos^2(Y)\rangle_\Omega-\frac{\Omega \tilde\psi_1\langle1\rangle_\Omega-\tilde{\psi}_1\langle\cos(Y)\rangle_\Omega}{\langle1\rangle_\Omega} \langle\cos(Y)\rangle_\Omega\right) \\&= 2\tilde{\psi}_1\left( \langle\cos^2(Y)\rangle_\Omega-\frac{\langle\cos(Y)\rangle_\Omega^2}{\langle1\rangle_\Omega}\right).
\end{split}
\end{equation}
After consolidating terms, the flow-modified Rutherford equation is
\begin{equation}\label{Appendix eq:Matched J, Mach By exp}
    \begin{split}
    \Delta' - 4\mu z_{\epsilon} &= \frac{4S}{ \tilde{\psi}_1^{1/2}} \frac{d\tilde{\psi}_1}{dt} A 
    +8 \tilde{\psi}_1^{1/2} A
    \\& + \frac{8}{\tilde \psi_1^{1/2}} \int^{1}_{\Omega_-}d\Omega\frac{M(\tilde\psi_1\Omega) M'(\tilde\psi_1\Omega)}{(1-M^2(\tilde\psi_1\Omega))}\left( \langle\cos^2(Y)\rangle_\Omega-\frac{\langle\cos(Y)\rangle_\Omega^2}{\langle1\rangle_\Omega}\right),
\end{split}
\end{equation}

\nocite{*}

\bibliography{bib}

\end{document}